\documentstyle[11pt]{article}

\begin{document}
\begin{flushright}
\today\\
IPM/P--99/7
\end{flushright}
\vspace{1 cm}
\begin{center}
{\Large \bf A MODEL UNIVERSE WITH VARIABLE SPACE }\\
{\Large \bf DIMENSION:}\\
{\Large \bf ITS DYNAMICS AND WAVE FUNCTION}\\
\vspace{1 cm}
{\it \large R. Mansouri \footnote{e-mail:mansouri@netware2.ipm.ac.ir}, 
F. Nasseri \footnote{e-mail:naseri@netware2.ipm.ac.ir}}\\
\vspace{3 mm}
{\it \large Department of Physics, Sharif University of Technology, P.O.Box
11365--9161, Tehran, Iran}\\
{\large and}\\
{\it \large Institute for Studies in Theoretical Physics and Mathematics,
P.O.Box 19395--5531, Tehran, Iran}\\
\end{center}
\begin{abstract}
Assuming the space dimension is not constant, but varies with the
expansion of the universe, a Lagrangian formulation of a toy universe
model is given. After a critical review of previous works, the field
equations are derived and discussed. It is shown that this generalization
of the FRW cosmology is not unique. There is a free parameter in the theory,
$C$, with which we can fix the dimension of space say at the Planck time.
Different possibilities for this dimension are discussed.
The standard FRW model corresponds to the limiting case $C \to +\infty$.
Depending on the
free parameter of the theory, $C$, the expansion of the model can behave
differently to the standard cosmological models with constant dimension.
This is explicitly studied in the framework of quantum cosmology.
 The Wheeler--De Witt equation is written down. It turns out that in our
 model universe, the potential of the Wheeler--DeWitt equation has different
 characteristics relative to the 
 potential of the de Sitter minisuperspace. Using the
 appropriate boundary conditions and the semiclassical approximation, we
 calculate the wave function of our model universe. In the limit of
$C \to +\infty$, corresponding to the case of constant space dimension, our
wave function has not a unique behavior. It can either leads to the
Hartle--Hawking wave function or to a modified Linde wave function, or to
a more general one, but not to that of Vilenkin. We also calculate the
probability density in our model universe. It is always more than the
probability density of the de Sitter minisuperspace in 3--space as suggested
by Vilenkin, Linde, and others. In the limit of constant space dimension,
the probability density of our model universe approaches to Vilenkin and
Linde probability density being $\exp(-2|S_E|)$, where $S_E$ is the Euclidean
action. Our model universe indicates therefore that the Vilenkin wave
function is not stable with respect to the variation of space dimension.
\end{abstract}
PACS: 98.80.Hw, 98.80.Bp, 04.60.Kz, 04.50.+h 
\section{INTRODUCTION} \label{sec-introduction}
Kaluza--Klein and string theories are well known for allowing 
the space dimension to be other than 3, but still an integer 
and a constant \cite{{tkal}, {mjdu}}. 
This, being considered for the high energy limit in the universe or for the 
dimension of space at the Planck time, has encouraged people to suggest that 
the dimension of space in the lower energy limit, or for the actual 
structured universe, is other than 3. 
Recently models have been proposed in which the universe has 
$(1+3+n)$--dimensional space
with Planck scale near the weak scale, with $n \geq 2$ new 
sub--millimeter sized dimensions \cite{N}.\\
In the last years there has been 
attempts to identify a fractal dimension for the matter distribution in  
space using either the cosmic microwave background radiation or the galaxy
distributions \cite{{phco}, {dnsc}}. 
Aside from the actual dimension of space or the matter 
distribution in it, it is interesting to study the cosmological consequences 
of a fractal and variable space dimension. All critiques of space 
dimensionality other than 3 rely upon cosmologically small scale 
observations \cite{mteg}. Therefore, one could ask about the consequences of 
a dynamical space dimension in cosmological time and space scales. A proposed 
way of handling such a concept is using the idea of decrumpling coming from 
the polymer physics [7--9].
The evolusion of the fractal dimension of a self--similar universe in the 
context of Newtonian gravitation is studied in \cite{9}.\\
Our toy model universe consists of small cells having arbitrary space    
dimensions. The effective spatial dimension of the universe in large
will depend on the configuration of the cells. As an example, take a
limited number of small three dimensional beads. Depending on how these
beads are embedded in space they can configure to a one-dimensional string, 
2-dimensional sheet, or a three dimensional sphere. This is the picture we
are familiar from the concept of crumpling in the polymer physics where a 
crumpled polymer has a dimension more than 1. Or take the picture of a 
clay which can be like a 3 dimensional sphere, or 
a 2 dimensional sheet, or even a one dimensional string; a picture we are 
familiar from the theory of the fluid membranes. In this picture the 
universe can have any space dimension. As it expands the space dimension
decreases continously.\\
In this paper we modify and study throughly the model proposed in
[11--13]. The original decrumpling model of the universe seems 
to be singularity free having two turning points for the space 
dimension \cite{{M1}, {M2}}. 
Lima {\it {et al}}. \cite{M3} criticize the way of 
generalizing the standard cosmological model to arbitrary space dimension 
used in \cite{M1} and propose another way of writing the field equations. 
Their model shows no upper bound for the dimension of space.
As the universe expands the spatial dimension decreases to $D=1$ \cite{M3}.\\
Later on this scenario was extended to the class of multidimensional 
cosmological models, where extra factor spaces play the role of the matter 
fields. In this multidimensional cosmological model an inflationary
solution was found together with the prediction that the universe starts 
from a nonsingular space time \cite{M4}. Melnikov {\it {et al}}. 
has written the Wheeler--DeWitt equation for multidimensional
cosmological model in any constant spatial dimension \cite{14}, see also 
\cite{15} and references therein.\\
We review briefly the previous works on the decrumpling models and show 
some difficulties in their formulations and conclusions. To remove these 
difficulties, we propose a new way to generalize the gravitational action
in constant dimension to the case of dynamical dimension. We use the 
Hawking--Ellis action for a prefect fluid in 3--space \cite{swha} and 
generalize it to the case of dynamical dimension. As we will show in details
the generalization of the gravitational action to the dynamical dimension is 
not unique. Moreover, in contrast to the earlier works 
in [11--13], we take into account the dependence of the 
measure of the action on space dimension. 
The generalization of the action, the Lagrangian, and the 
equations of motion to dynamical space dimension is then done in two ways.
Studing the time evolution of the spatial dimension, we obtain some numerical 
results for the turning points in our model. \\ 
In the second part of the paper we study the quantum cosmology of our model 
universe.  The Wheeler--DeWitt equation is written and the zero points of its
potential are discussed. It is worth noticing that the 
potential of our model has completely different behavior from the 
potential of the de Sitter minisuperspace in 3--space.
Imposing the appropriate boundary condition in the limit $ a \to +\infty$, and 
using the semiclassical approximation, we obtain the wave function 
of our model. It is then seen that in the limit of constant space dimension,
our wave function is not well--defined. 
We show that it can 
approach to the Hartle--Hawking wave function 
or to the modified Linde wave function, but not to that of Vilenkin.
We also estimate 
the probability density in our model universe. In the limit 
of constant spatial dimension, it approaches to Vilenkin, Linde and others'
proposal \cite{17}; i.e. to the probability density 
${\cal{P}} \propto \exp(2S_E)$, or more generally $\exp(-2|S_E|)$, where
$S_E$ is the Euclidean action of the classical instanton solution.\\
The paper is organized as follows. In Sec.~\ref{sec-cellular} we discuss  
the dimensional constraint, proposed in ~\cite{M1}. In 
Sec.~\ref{sec-review}, we briefly review the earlier works on the decrumpling 
model and discuss their shortcomings. In Sec.~\ref{sec-new}, a new 
way to generalize the gravitational-- and the matter--action to the case of
dynamical dimension is presented. The equation of motion and the time evolution 
equation of spatial dimension is then obtained. Sec.~\ref{sec-wave} is 
devoted to the Wheeler--DeWitt equation of our model and the calculation of 
its wave function in semiclassical approximation. The probability density in 
our model universe is then estimated semiclassically.
Section~\ref{sec-concluding} concludes the discussion of our toy model 
universe. 

\section{MOTIVATION AND CONSTRAINT OF THE MODEL} \label{sec-cellular}
 
There are observational evidences that the matter distribution in the 
universe, up to the present observed limits of $100 h^{-1} {\rm Mpc}$, is a 
fractal or multifractal \cite{phco,dnsc} having a dimension of about 2. 
This should be considered as an effective dimension of space without
questioning the three dimensionality at cosmologically small 
scales \cite{mteg}. \\
To interpret the effective space dimension $D$ we follow the picture proposed 
in \cite{M1}--\cite{M3}. Imagine the fundamental building blocks of the universe 
as cells being arbitrary dimensional and having, in each dimension, a 
characteristic size $\delta$ which maybe of the order of the Planck 
length ${\cal O} (10^{-33} {\rm cm})$ or even smaller. These ``space--cells'' are 
embedded in a $\cal D$ space, where $\cal D$ may be up to infinity. Therefore, 
the space dimension of the universe depends on how these fundamental cells 
are configured in this embedding space. The universe may have begun from a 
very crumpled state having a very high dimension $\cal D$ and a size 
$\delta$, then have lost dimension through a uniform decrumpling which we see 
it as a uniform expansion. The expansion of space, being now understood as a 
decrumpling of cosmic space, reduce the space--time dimension continuously
from ${\cal {D}} + 1$ to the present value $D_0 + 1$. We do not fix $D_0$ to 
allow being other than 3.\\ 
Here, we are interested in the correlation between the radius of our model 
universe and its dimension. In the next section, we will implement this idea 
in the Lagrangian formulation. The first major formal difficulty in 
formulating a space--time theory with variable space dimension is that the 
measure of the integral action is variable and therefore some part of
integrand. The variational calculus for such a case has not been
formulated yet.
We are, however, interested in a cosmological model for the actual universe. 
Therefore, it is reasonable to accept the cosmological principle: the 
homogeneity and isotropy of space. This simplifies the matter 
substantially, so that it becomes possible to formulate a Lagrangian for 
the theory and write the corresponding field equations. 

\subsection{Relation between the effective space dimension $D(t)$ and 
characteristic size of the universe $a(t)$} 
\label{subsec-relation}

Assume the universe consists of a fixed number $N$ of universal cells having
a characteristic length $\delta$ in each of their dimensions. The volume of 
the universe at the time $t$ depends on the configuration of the cells. It is 
easily seen that \cite{{M1}, {M2}}
\begin{equation}
\label{vol}
{\rm vol}_D({\rm cell})={\rm vol}_{D_0}({\rm cell}) \delta^{D-D_0}.
\end{equation}
Interpreting the radius of the universe, $a$, as the radius of gyration of 
crumpled ``universal surface'' \cite{pjde}, the volume of space can be written 
\cite{{M1}, {M2}}:  
\begin{eqnarray}
a^D&=&N {\rm vol}_D({\rm cell})\nonumber\\
   &&=N {\rm vol}_{D_0}({\rm cell})\delta^{D-D_0}\nonumber\\
   &&={a_0}^{D_0} \delta^{D-D_0},\nonumber\\
\end{eqnarray}
or 
\begin{equation}
\label{con}
(\frac{a}{\delta})^D=e^C,
\end{equation}
where $C$ is a universal positive constant. Its value has a strong influence 
on the dynamics of space--time, for example on the dimension of space say 
at the Planck time. Hence, it has physical and cosmological consequences and 
may be determined by observation. We coin the above relation as ``dimensional 
constraint'' which relates the ``scale factor'' of our model universe to the 
space dimension. Aside from any dynamics, which we will discuss in the next 
sections, it is worth discussing the dimensional constraint thoroughly.

\subsection{Discussion on the dimensional constraint}
\label{subsec-discussion}

Eq.(\ref{con}) can be written as
\begin{equation}
\label{c1}
\frac{1}{D}=\frac{1}{C} {\ln \frac{a}{a_0}}+\frac{1}{D_0}.
\end{equation}
The time derivative of the above relation leads to 
\begin{equation}
\label{c2}
\dot {D}=-\frac{D^2}{C} \frac{\dot a}{a}.
\end{equation}
In the above relation $a_0$ is the present scale factor of the universe 
corresponding to the space dimension $D_0$. The values of $C$ and $\delta$ 
can be calculated now in terms of other known quantities:
\begin{equation}
\label{c3}
C=\frac{D D_0}{(D-D_0)} \ln (\frac{H_0^{-1}}{a}),
\end{equation}
and
\begin{equation}
\label{c4}
\log(\frac{\delta}{a})=\frac{\log {\frac{a}{H_0^{-1}}}}{-1+
\frac{D}{D_0}},
\end{equation}
where we have replaced $a_0$ by the present value of the Hubble radius 
${H_0}^{-1} = 3000 h^{-1} {\rm Mpc} = 9.2503\times 10^{27} h^{-1} {\rm cm}$. As the 
values of $C$ and $\delta$ are not very sensitive to $0.5 <h <1$ we take for   
simplicity $h = 1$.\\
Note that for $D \rightarrow  +\infty$ the radius of the universe, $a(t)$,
tends to $\delta$ (see Eq.\ref{c4}). In principle, $\delta$ may be greater 
or less than the Planck length $l_{Pl}$. Assuming $\delta = l_{Pl} = 
1.6160 \times 10^{-33} {\rm cm}$ and taking 
$D_0=3$ we obtain from (\ref{c3}) $C = 419.7$. Values of $\delta$, or the 
minimum radius of the universe, less than the Planck length 
correspond to $C>419.7$. Assuming $C < 419.7,$ we obtain for the minimum 
radius of the universe, $\delta$, values bigger than the Planck length.
Not all dimensions of space at the present or the Planck time are 
of interest. We are just interested in the experimental value $D_0 = 3-(5.3
{\pm}2.5)\times 10^{-7}$ \cite{azme} based on the $g-2$ factor of electron, 
or ${D_0} \simeq 2$ as fractal dimension for matter distribution in the 
universe coming from cosmological considerations \cite{phco}. 
At the Planck time we assume the space dimension, $D_{Pl}$, to be about 3 
or one of the values 4, 10, or 25 coming from supergravity \cite{mjdu} and 
superstring theories \cite{jhsc,mbgr}. It is reasonable to assume that the 
minimum radius of the universe is less than the Planck length. Therefore, in 
the following we will assume $C>419.7$. Note that, in an expanding 
universe, the positivity of $C$ gives the decreasing of the spatial dimension
in the course of time (cf. Eq.\ref{c2}). Table I shows values of $C$ and 
$\delta$ for different values of $D_{Pl}$ assuming $D_0 \simeq 3$. Similarly, 
table II shows the same values for $D_0 = 2$.\\
Figs. 1--4 demonstrate changes of $C$ or $\delta$ vs $D_{Pl}$ and $D_0$.
Figures 1--2 demonstrate the changes of $C$ vs $D_{Pl}$ (or $\log D_{Pl}$)
for different values
of $D_0$ in different ranges. In order to illustrate the behavior of $\delta$ 
vs $D_{Pl}$, Figs. 3 and 4 are drawn. In general, for a fixed value of $D_0$,
when $C$ increases $\delta$ and $D_{Pl}$ decrease. \\
\small{
TABLE I. Values of $C$ and $\delta$ for some interesting 
values of $D_{Pl}$ assuming $D_0 \simeq 3$. 
The reason for taking $D_0=3-5.3\times10^{-7}$ is based on the experimental 
data by measuring g--factor of electron (for details see Ref.\cite{azme}). 
$D_{Pl}$ could be any value, here we take an arbitrary 
value for it, of course with condition $D_{Pl} \geq D_0$. 
}
\begin{center} 
\begin{tabular}{|p{2.5 cm}|p{2.0 cm}|p{2.5 cm}|p{3.0 cm}|}  \hline\hline
$D_0$ & $D_{Pl}$ & $C$ & $\delta ({\rm cm})$ \\ \hline
 $3-5.3\times10^{-7}$ & $3+10^{-3}$ & $1.2588\times10^6$ & $4.4747\times10^{-182210}$ \\  \hline
 $3$ & $4$ & $1.6788{\times}10^{3}$ & $8.6158\times 10^{-216}$ \\  \hline
 $3$ & $10$ & $5.9957\times10^2$ & $1.4771\times 10^{-59}$ \\  \hline
 $3$ & $25$ & $4.7693\times10^2$ & $8.3811\times 10^{-42}$ \\  \hline                
 $3$ & $+\infty$ & $4.1970\times 10^2$ & $1.6160\times 10^{-33}$ \\  \hline\hline
\end{tabular}
\end{center}
\vspace{0.25 cm}  
\begin{center}
\small{
TABLE II. Values of $C$ and $\delta$ for different values 
of $D_{Pl}$ assuming $D_0=2$.
}
\begin{tabular}{|p{1.5 cm}|p{2.5 cm}|p{3.0 cm}|}  \hline\hline
$D_{Pl}$ & $C$ & $\delta ({\rm cm})$ \\ \hline
 $4$ & $5.5960 \times 10^2$ & $2.8231 \times 10^{-94}$ \\ \hline 
 $10$ & $3.4975 \times 10^2$ & $1.0447 \times 10^{-48}$  \\ \hline 
 $25$ & $3.0413 \times 10^2$ & $8.4171 \times 10^{-39}$ \\ \hline
 $+\infty$ & $2.7980 \times 10^2$ & $1.6160 \times 10^{-33}$ \\ \hline\hline
\end{tabular}
\end{center}
\section {REVIEW OF THE DECRUMPLING UNIVERSE MODEL}
\label{sec-review}
The assumption of variable space dimension brings in serious difficulties
in formulating field equations. To formulate a gravitational theory based 
on geometry we do need a tensor calculus which is not defined for 
dynamical space dimension.
It may be possible to write an action but we do not know how to 
formulate a calculus of variation in the case that the measure of the 
integral is itself a dynamical variable. 
Therefore, we have to look for methods to 
define or obtain the field equations. The homogeneity and isotropy of 
space--time, being the main characteristics of the Friedmann models, allow 
us to formulate an action and a variational method to obtain the field 
equations. Although this do not leads to a unique model for a universe having 
a continuously varying dimension, it leads to interesting results 
and maybe the simplest way to implement the idea of having a space with 
variable dimension.

\subsection{Action and Lagrangian}
\label{subsec-action}
As was mentioned before, we have no way of defining an action for a general 
metric implementing the dynamical space dimension. 
Therefore, let us define it first for the special FRW metric in an 
arbitrary fixed space dimension $D$ and then try to generalize it to variable
dimension. Now, take the metric in constant $D+1$ dimension in the 
following form (we use natural units or high energy physics units  
in which the fundamental constants are $\hbar=c=k_B=1,\, 
G={l_{Pl}}^2=1/{m_{Pl}}^2$).
\begin{equation}
\label{III.1}
ds^2=-N^2(t)dt^2+a^2(t) {d \Sigma_k} ^2,
\end{equation}
where $N(t)$ denotes the lapse function and ${d \Sigma_k}^2$ is the line element 
for a D--manifold of constant curvature $k=+1,0,-1$, corresponding
to the closed, flat, and hyperbolic spacelike sections, respectively. 
The Ricci scalar is given by 
\begin{equation}
\label{III.1a}
R=\frac{D}{N^2}[\frac{2 \ddot a}{a}+(D-1) ((\frac{\dot a}{a})^2+
\frac{N^2 k}{a^2}) - \frac{2 \dot a \dot N}{a N}].
\end{equation}
For simplicity, let us now take $k=0$ in our treatments as done in 
Refs.~\cite{{M1}, {M3}}.
Substituting from Eq.(\ref{III.1a}) in the Einstein--Hilbert action for pure 
gravity,
\begin{equation}
\label{III.2}
S_{G}= \frac{1}{2 \kappa} \int{d^{(1+D)} x \sqrt{-g} R},
\end{equation}
where $\kappa = 8\pi G$, we are led to  
\begin{eqnarray}
\label{III.4}
S_{G} &=& \frac{1}{2 \kappa } \int d^{(1+D)} x  \{ 2D \frac{d}{dt}
          (\frac{\dot a}{aN} a^D)-\frac{D(D-1)}{N}(\frac{\dot a}{a})^2 a^D \}  \nonumber\\
      &&= \frac{1}{2 \kappa} \int d^{(1+D)} x \ \{{\rm total} \ {\rm time}\ {\rm derivative} \nonumber\\
      &&- \frac{D(D-1)}{N}(\frac{\dot a}{a})^2 a^D \}.
      \end{eqnarray}
The Lagrangian then becomes
\begin{equation}
\label{III.5}
L^{(0)}_G=-\frac{D(D-1)}{2 \kappa N}(\frac{\dot a}{a})^2 a^D.
\end{equation}
Following Ref.~\cite{M1}, we introduce the matter Lagrangian for a prefect fluid as
\begin{equation}
\label{III.6}
L^{(0)}_M=-\frac{\tilde \rho N^2}{2}+\frac{{\tilde p} D a^2}{2},
\end{equation}
where 
\begin{eqnarray}
\tilde \rho  &:=& \frac{\rho}{N} a^D,   \\
\tilde p    &:=& p N a^{D-2} \delta^{ij},  
\end{eqnarray}
and $\rho$ and $p$ being the energy density and pressure, respectively. 
Therefore, the complete Lagrangian is 
\begin{equation}
\label{III.6b}
L^{(0)}=-\frac{D(D-1)}{2 \kappa N}(\frac{\dot a}{a})^2 a^D+(-\frac
{\tilde \rho N^2}{2}+\frac{{\tilde p} D a^2}{2}).  
\end{equation}
This Lagrangian suffers from the fact that its dimension is not 
(length)$^{-1}$ as we know it from the $D = 3$ case, 
and depends on the space dimension.
To remedy this shortcoming which is an obstacle for generalizing to
variable dimension, we first assume that the dimension of $\kappa$ is 
constant and equal to (length)$^{D_0-1}$ for $D_0 = 3$. This is equivalent to the 
assumption that for $\kappa$ we take the usual three dimensional
gravitational constant. We then multiply the above Lagrangian by 
${a_0}^{D_0-D}$, where $a_0$ is the present  scale factor of the 
universe. For simplicity, we have assumed that $a$ and $a_0$ have the  
dimension of length. Omitting the constant factor ${a_0}^{D_0}$ we arrive at  
\begin{equation}
\label{III.7}
{\cal L}=-\frac{D(D-1)}{2\kappa N}(\frac{\dot{a}}{a})^2
(\frac{a}{a_0})^D+(-\frac {\hat{\rho} N^2}{2}+\frac{\hat{p}Da^2 }{2}),
\end{equation}
where
\begin{eqnarray}
\hat{\rho}=\frac{\rho}{N} (\frac{a}{a_0})^D,\;\;\;\;\\
\hat{p}=pa^{-2}N(\frac{a}{a_0})^D.
\end{eqnarray}
This Lagrangian has the dimension (length)$^{-4}$ which is 
due to omitting of the factor ${a_0}^{D_0}$ for $D_0=3$. 
It is now in a form ready to generalize to any variable dimension.
We consider the gravitational coupling constant $\kappa$ as a constant 
throughout this paper. The dependence of $\kappa$ on the constant spatial 
dimension is studied, in Ref.~\cite{21}.

\subsection{Field equations}
\label{subsec-field}
For $D=D_0$, the Friedmann equations derived by the Einstein field equations
are
\begin{equation}
\label{III.8}
\frac{1}{N^2}(\frac{\dot a}{a})^2=\frac{2 \kappa \rho}{D_0 (D_0-1)},
\end{equation}
and
\begin{equation}
\label{III.9}
\frac{\ddot a}{N^2 a}=\frac{\kappa}{D_0(D_0-1)}[(2-D_0) \rho -D_0 p],   
\end{equation}
with $k = 0$. Note that the field equations  
are written in the gauge $\dot N = 0$.
Eqs.(\ref{III.8}, \ref{III.9}) are the $00$-- and non--vanishing $ij$--component 
of the Einstein field equations, respectively. There is an alternative 
Lagrangian method of deriving these field equations which can be 
generalized to the case of dynamical space dimension. Consider first the 
usual case of $D=D_0={\rm const}.$ It is easy to show that (\ref{III.8}) 
is the equation of motion corresponding to variation of the Lagrangian 
(\ref{III.7}) with respect to $N$. Moreover, variation of (\ref{III.7}) 
with respect to $a$ yields 
\begin{equation}
\label{III.10}
-(D_0-1)[\frac{\ddot a}{a}+(\frac {D_0-2}{2})(\frac{\dot a}{a})^2]=
\kappa p N^2,
\end{equation}
which is a combination of (\ref{III.8} and \ref{III.9}).
Note that in varying the Lagrangian with respect to $N$ and $a$ the quantities 
$\hat \rho$ and $\hat p$ are considered as constants \cite{M1}--\cite{M3}.
Therefore, for constant space dimensions this Lagrangian method leads
to the familiar Friedman equations derived by the Einstein field equations.
The continuity equation for prefect fluid is obtained from the Friedman
equations (\ref{III.8} and \ref{III.9}) or (\ref{III.8} and \ref{III.10}):
\begin{equation}
\label{III.11}
\frac{d}{dt}(\rho a^{D_0})+p \frac{d}{dt} (a^{D_0})=0.
\end{equation}
This, in addition to an equation of state, completes the field equations 
for a flat FRW universe for any fixed space dimension $D_0$. \\
To generalize the cosmological model to spaces with variable dimension, 
we are in principle free to choose any of the above approaches.
But, it is worth noticing that the tensor calculus is meaningless in 
the case of dynamical dimension, since the index of a tensor is always 
constant and integer number. Therefore, there is no way of generalizing the 
Einstein equations directly. In contrast, it is possible to write down
generalized 
Lagrangian, such as (\ref{III.7}), due to the fact that space is homogeneous
and the dynamical part of the measure is well defined and separated.
Hence, we are left with the second approach as the only possible way of 
formulating a cosmological model incorporating the dynamical nature of 
space dimension. However, we are faced now with different kind of 
non--uniqueness in formulating the Lagrangian or the field equations, 
which are discussed in the following sections.

\subsubsection{Original decrumpling model}

Following \cite{M1}, we vary the Lagrangian (\ref{III.7}) with respect to  
$a$, assuming the gauge $\dot N = 0$ and taking into account the relation 
(\ref{con}):
\begin{eqnarray}
\label{III.12}
\frac{(D-1)}{N} \{\frac{\ddot{a}}{a}+[\frac{D^2}{2D_0}-1-\frac{D(2D-1)}
{2C(D-1)}](\frac{\dot{a}}{a})^2\}\nonumber\\
+N \kappa p (1-\frac{D}{2C})=0.
\end{eqnarray}
It is easily seen that this equation leads to the field equation 
(\ref{III.10}) for $C \rightarrow  +\infty$ and $D=D_0 = {\rm const}$.
Authors of \cite{M1} prefer to take the continuity equation as the second 
field equation. By a dimensional reasoning the continuity equation 
(\ref{III.11}) is generalized to \cite{M1}
\begin{equation}
\label{III.13}
\frac{d}{dt}(\rho (\frac{a}{a_0})^{D})+p \frac{d}{dt} (\frac{a}{a_0})^{D}=0.
\end{equation}
Now, a qualitative behavior of this toy model is obtained via a first 
integral of motion. From the Lagrangian (\ref{III.7}), the Hamiltonian 
is defined as  
\begin{equation}
\label{III.14}
{\cal H} =  -\frac{(D-1)}{2 \kappa} \frac{C^2 
{\dot D}^2}{D^3}e^{C} (\frac{\delta}{a_0})^{D} -
(\frac{\hat p}{2} D \delta^2 e^{2C/D} - 
 \frac{\hat {\rho} N^2}{2}),
\end{equation}
with
\begin{equation}
\label{III.15}
\frac{ d {\cal H}}{d t}=-\frac {\partial {\cal L}}{\partial t},
\end{equation}
and taking $\hat \rho$ and $\hat p$ as source [1-3]:
\begin{eqnarray}
\label{III.15a}
&&\frac{d}{dt}[-\frac{C^2}{2 \kappa N} e^C (\frac{\delta}{a_0})^D 
\frac{D-1}{D^3} {\dot D}^2-\frac{\hat p}{2} D \delta ^2 e^{2C/D}]\nonumber\\
&&=-\frac{D}{2} 
\delta^2 e^{2C/D} \frac{d \hat p}{d t},
\end{eqnarray}
which leads to
\begin{eqnarray}
\label{III.16}
\frac{C^2}{2 \kappa N} \frac{d}{dt}(\frac{D-1}{D^3} e^{-\frac{CD}{D_0}}
{\dot D}^2)+p \dot D e^{-\frac{CD}{D_0}}(\frac{1}{2}-\frac{C}{D})=0.
\end{eqnarray}
Taking the derivatives with respect to $D$ instead of $t$, we arrive at  
\begin{eqnarray}
\label{III.17}
\frac{C^2}{2 \kappa N} \frac{d}{dD}(\frac{D-1}{D^3} e^{-\frac{CD}{D_0}}
{\dot D}^2)+p e^{\frac{-CD}{D_0}}(\frac{1}{2}-\frac{C}{D})=0,
\end{eqnarray}
which is equivalent to 
\begin{eqnarray}
\label{III.17a}
\frac{C^2}{2 \kappa N}(\frac{D-1}{D^3} e^{-\frac{CD}{D_0}}
{\dot D}^2)+ \int ^{D}_{D_0} dD' p(D') e^{-\frac{CD'}{D_0}}\nonumber\\
\times (\frac{1}{2}-\frac{C}{D'})=0.
\end{eqnarray}
This is interpreted now as the equation of motion for $D$, having a 
vanishing total energy, in the potential
\begin{eqnarray}
\label{III.18}
{\cal U}(D) :=\int ^{D}_{D_0} dD' {p(D') e^{-\frac{CD'}{D_0}}
(\frac{1}{2}-\frac{C}{D'})}.
\end{eqnarray}
The kinetic energy term is given by
\begin{eqnarray}
\label{III.19}
{\cal T} := \frac{C^2}{2 \kappa N}(\frac{D-1}{D^3} e^{-\frac{CD}{D_0}}
{\dot D}^2).
\end{eqnarray}
The matter content is taken in the form of radiation with the following 
equation of state 
\begin{equation}
\label{III.20}
p=\frac{\rho}{D}.
\end{equation}
The kinetic energy is always positive for $D > 1$. Therefore the potential 
energy has to be negative for real dimensions. There is a minimum at
$D=2C$ for the potential (\ref{III.18}) where it is negative. 
Its behavior at large $D$ is given by
\begin{equation}
{\cal{U}}(D) \simeq \frac{D_0}{2C} e^{-C} (\frac{D}{D_0})^{\frac{C}{D_0}},
\end{equation}
and for $D$ near zero, assuming the pressure remains finite (nonzero)  
\begin{equation}
{\cal{U}}(D) \simeq -C \ln D.
\end{equation}
Therefore, $\cal U$ tends to infinity when $D \rightarrow +\infty$
as well as $D \rightarrow 0$. This means that there are two turning points
for $D$ (where $\dot D=0$), one above $D=2C$ and the other below it. Appropriate 
assumptions about the matter content of the universe would bring the lower 
turning point to be $D > 1$. This leads to a model universe without any 
singularity. Even dimension of the model, being a dynamical quantity remains 
finite. The behavior of the spatial dimension with respect to the time near 
$D = D_0 = 3$ can be obtained by expanding (\ref{III.17a}) using 
$D=D_0 + \epsilon(t)$. Assuming
\begin{equation}
\label{III.22}
\frac{C \epsilon}{D_0} \ll 1,
\end{equation}
they obtain to the lowest order in $\epsilon$ 
\begin{equation}
\label{III.23}
{\dot \epsilon}^2=\frac{2 \kappa N {D_0}^2 p_0}{C(D_0-1)} {\epsilon},
\end{equation}
which has the solution
\begin{equation}
\label{III.24}
\epsilon = \frac{ \kappa N {D_0}^2 p_0}{2C(D_0-1)} (t_0 - t)^2,
\end{equation}
where $t_0$ is the time corresponding to $D_0$. Using this relation, it 
has then been shown that for $\epsilon (t_0)$ of the order of $10^{-6}$ 
the relative change of space dimension is just about two order of 
magnitude within a time range of about 10 times the present accepted age 
of the universe \cite{M1}--\cite{M3}. 
That is, as we go backward in time there will be almost    
no change in the space dimension as far as 100 billion years.\\ 
It should be noted, however, that the change of variable from $t$ to
$D$ in (\ref{III.17}) is not allowed (see Sec.~\ref{subsubsec-incorrect}). 
Therefore, the result we just mentioned are questionable. 
Besides this, the authors of [11--13] in the course of the above 
calculation use the value $C = 600$, which is not consistence with 
assuming $D_{Pl} \simeq 3$. For this to be the case we 
must have $C \simeq 10^6$, as can be seen from Table I. 

\subsubsection{Modification of the original decrumpling model}

Lima {\it {et al}}. \cite{M3} propose an alternative way of writing the field 
equations. They bring in the lapse function and consider the variation of the  
Lagrangian with respect to it as the second field equation. This corresponds
to the $00$--component of the Einstein field equations (cf. Eq.(\ref{III.8}) 
for constant dimension): 
\begin{eqnarray}
\label{III.32}
\frac{1}{N^2} (\frac{\dot a}{a})^2= \frac{ 2 \kappa \rho}{D(D-1)}.
\end{eqnarray}
The two field equations (\ref{III.12}) and (\ref{III.32}) are consistent with the 
continuity equation (\ref{III.13}). Now, assuming a radiation dominated universe 
(i.e. $p=\rho /D$), Eqs.(\ref{c2}, \ref{III.13}, and \ref{III.32}) leads 
to the following evolution equation for the spatial dimension $D$
\begin{equation}
\label{III.33}
{\dot D}^2=\frac{AD^3 e^{C(\frac{D}{D_0}-1)}(\frac{D}{D_0})^{\frac{C}{D_0}}}
{C^2(D-1)},
\end{equation}
where $A=2\kappa \rho_0 N^2$ and $\rho_0$ is the energy density of the 
universe corresponding to $D=D_0$. This ``energy equation'' for $D$  
corresponds now to (\ref{III.17a}). However, there is an 
inconsistency between them (see Sec.~\ref{subsubsec-incorrect}). 
For $D<1$ the kinetic term is again negative. 
Now, the potential energy is always negative for  
$1 < D < +\infty$. Thus there is no turning point in this modified model.

\subsection{Critique of the original decrumpling Model and its modification}
\label{subsec-critique}
As explained before, there is not a unique formulation for the 
cosmological models with variable space dimension, and no principle from 
which we could derive unique field equations. Therefore, the observational 
consequences are the only way of looking for validity of the models. 
Despite this non--uniqueness and possible observational viability,
we would like to mention some theoretical deficiencies of 
the above models.

\subsubsection{Measure of Action}
\label{subsubsec-measure}
The main difficulty in formulating a variational calculus for a theory  
with variable space dimension is the dependence of the measure of action 
on dimension, making the measure a dynamical quantity. As we will see 
in the next section, this dependence seriously influences the field equations.
Models discussed up to now \cite{{M1}, {M3}} do not consider this dependence. 
Taking into account the  measure of the integral will also change the 
equation of continuity (see Eq.\ref{IV.20a}).

\subsubsection{Incorrect time evolution equation of the spatial dimension}
\label{subsubsec-incorrect}
We have seen that two equations (\ref{III.17a}) and (\ref{III.33}) leads to 
quite different results regarding the behavior of $D$. 
Lima {\it {et al}}. \cite{M3} are able to use
actually one of the field equations directly, i.e. (\ref{III.32}). Authors
of Ref.~\cite{M1}, however, use equation (\ref{III.17a}), which is based 
on a non-valid change of variables in going from Eq.(\ref{III.16}) to 
(\ref{III.17}). Therefore, their equation for the dynamics of $D$, i.e. 
(\ref{III.17a}), is not valid.\\
The point is better understood if we write the Hamiltonian in its correct
canonical form. Substituting the momentum conjugate to the space dimension,
$p_D$, in (\ref{III.14}), we obtain for the Hamiltonian
\begin{eqnarray}
\label{III.41}
{\cal {H}}({p_D}, {D})&=&-\frac{\kappa N D^3 e^{-C(1-\frac{D}{D_0})}}
{2 C^2 (D-1)} {p_D}^2+\frac{1}{2}(\hat \rho N^2\nonumber\\ 
&&- \hat p D \delta^2 e^{2C/D}).
\end{eqnarray}
Now, using ${\cal H}$ as a first integral of motion (see Eq.\ref{III.15}), we 
obtain
\begin{equation}
\label{III.43}
\frac{\kappa N}{2 C^2} 
\frac{d}{d t}(\frac{D^3 e^{CD/D_0}}{D-1} {p_D}^2)+
p e^{C(2- \frac{D}{D_0})} \dot D (\frac{1}{2}-\frac{C}{D})=0.
\end{equation}
Since $p_D$ and $D$ are independent variables, it is obvious that we are not 
allowed to change the differential variable from $t$ to $D$. Note that
Eq.(\ref{III.43}) is just Eq.(\ref{III.16}) where $p_D$ is substituted in 
terms of $\dot D$. Indeed, it is easy to show that Eq.(\ref{III.16}) or 
(\ref{III.43}) leads to (\ref{III.12}). 

\section{NEW MODELS FOR A UNIVERSE WITH VARIABLE SPACE DIMENSION}
\label{sec-new}

We have mentioned some of the shortcomings of the original model, regarding 
the field equations and its results. Now, as we have mentioned before even 
the Lagrangian is not unique. Here we try to remedy some of the shortcomings
and also write down other possible Lagrangians.\\
The matter part of the Lagrangian was written in a form such that $\hat \rho$ 
and $\hat p$ could be taken as constant in the variation of Lagrangian. 
Now, for the case of variable dimension this procedure may not work. Therfore, 
we prefer to use another procedure to write down the matter Lagrangian which 
is due to Hawking and Ellis \cite{swha}. 
This is outlined in section~\ref{subsec-hawking}. 
In section~\ref{subsec-new}, we formulate new  
actions, and Lagrangians, and then derive the corresponding field 
equations, using the new matter Lagrangian (\ref{IV.10}) and implementing 
the dynamic character of the measure. 

\subsection{Hawking--Ellis action of a prefect fluid in constant space 
dimension} \label{subsec-hawking}

We want to obtain the energy--momentum tensor of a prefect fluid in constant 
dimension, via an action formalism, in the form
\begin{equation}
\label{IV.1}
T^{\mu \nu}=(\rho + p) U^{\mu} U^{\nu}+pg^{\mu \nu},
\end{equation}
where $U$ is the timelike velocity four vector satisfying
\begin{equation}
\label{IV.2}  
g_{\alpha \beta} U^{\alpha} U^{\beta}=-1.
\end{equation}
The fluid can be described by a function ${\mu}$, called the density, and a 
congruence of timelike flow lines. The fluid current vector, defined by 
$j^{\alpha}=\mu U^{\alpha}$, is conserved:  
\begin{equation}
{j^{\alpha}}_{{;} {\alpha}}=0,
\end{equation}
where `` $;$ '' denotes covariant derivative.
Taking the elastic potential (or internal energy) ${\epsilon}$ as a function 
of ${\mu}$ the action is written as \cite{swha}
\begin{equation}
\label{IV.3}  
S_M=-\int d^{(D+1)}x {\sqrt{-g}}\; \mu \; (1+\epsilon).
\end{equation}
The calculations may be simplified by noting that the conservation of the 
current can be expressed as
\begin{equation}
\label{IV.4}  
{j^{\alpha}}_{{;}{\alpha}}=\frac{1}{\sqrt{-g}}\frac{ \partial}{\partial x^{\alpha}}
(\sqrt{-g} j^{\alpha})=0.
\end{equation}
Given the flow lines, the conservation equations determine $j^{\alpha}$ 
uniquely at each point on a flow line in terms of the initial values at some 
given point on the same flow line. Therefore, $(\sqrt{-g})j^{\alpha}$ is 
unchanged when the metric is varied. But 
\begin{equation}
\label{IV.5}  
{\mu} ^2=-g^{-1} ((\sqrt{-g}j^{\alpha})(\sqrt{-g}j^{\beta}))g_{{\alpha} {\beta}},
\end{equation}
so
\begin{equation}
\label{IV.6}  
2 \mu \delta \mu=(j^{\alpha} j_{\alpha} g^{\mu \nu}-j^{\mu}j^{\nu})\delta 
g_{\mu \nu}.
\end{equation}
Now, using the definition of energy--momentum tensor \cite{swei} as
\begin{equation}
\label{IV.7}  
\delta S_M=  \frac{1}{2} \int d^{(D+1)} x \sqrt {-g} T^{\mu \nu} 
\delta g_{\mu \nu},
\end{equation}
one gets Eq. (\ref{IV.1}) for the energy momentum--tensor of a perfect 
fluid in which
\begin{equation}
\label{IV.8}
\rho=\mu(1+\epsilon),
\end{equation}
and
\begin{equation}
\label{IV.9}
p=\mu^2(\frac
{d\epsilon}{d \mu}), 
\end{equation}
are the energy density and the pressure respectively. Hence the action for a 
prefect fluid may be expressed as
\begin{equation}
\label{IV.10}  
S_M=-\int d^{(D+1)}x \sqrt{-g} \;\rho.
\end{equation}
Taking a FRW metric given by Eq.(\ref{III.1}), it is easy to see that for a
comoving observer whose contravariant velocity four vector is 
\begin{equation}
\label{IV.11}
U^{\alpha}=(N^{-1},0,...,0),
\end{equation}
the energy--momentum tensor is given by 
\begin{equation}
\label{IV.12}
T^{\mu \nu}=diag( \rho N^{-2}, p a^{-2},...,pa^{-2}).
\end{equation}
In the variation of $S_M$ to obtain the energy--momentum tensor we have 
used the following relations: 
\begin{equation}                                                            
\label{IVa.12}                                                              
\delta \mu = - \frac{\mu D}{a} \delta a,\;\;\; {\rm if} \;\; a \to a+\delta a,    
\end{equation}                                                               
and                                                                         
\begin{equation}                                                            
\label{IVb.12}
\delta \mu=0, \;\;\; {\rm if} \;\; N \to N+\delta N.
\end{equation}
The matter action (\ref{IV.10}) is now in suitable form to be generalized 
to variable space dimension.

\subsection{Definition of our model universe and its field equations} 
\label{subsec-new}

We are now in a position to write down the complete action for matter and 
gravity in $D$ space dimension. Here our treatments are for each kind of 
D--dimensional topology (closed, flat, and open), in contrast to the earlier 
works in [11--13] which are for the special case of a flat FRW 
universe. Using the relations (\ref{III.1a}, \ref{III.2}, \ref{IV.10}) 
we obtain
\begin{eqnarray}
\label{IV.15a}
S_0 &:=& S_G + S_M \nonumber\\ 
&&=\int d^{(1+D)}x N a^D \{ \frac{D}{2 \kappa N^2}[\frac{2 \ddot a}{a}+(D-1)
((\frac{\dot a}{a})^2\nonumber\\
&&+\frac{N^2 k}{a^2})- \frac{2 \dot a \dot N}{aN}]-\rho \}.
\end{eqnarray}
Now, trying again to make the action dimensionless, we have to multiply it by 
the factor $a_0^{D_0-D}$ (cf. Sec.~\ref{subsec-action}). Omitting the constant part of this 
factor, ${a_0}^{D_0}$, and integrating over the space part of the action  
(\ref{IV.15a}), we reach to 
\begin{eqnarray}
\label{IV.15ab}
S&:=&S_{0} a^{-{D_0}}\nonumber\\
&&=\int dt V_D N (\frac{a}{a_0})^D \{ \frac{D}{2 \kappa N^2}
[\frac{2 \ddot a}{a}+(D-1)((\frac{\dot a}{a})^2\nonumber\\
&&+\frac{N^2 k}{a^2})- \frac{2 \dot a \dot N}{aN}]-\rho \},
\end{eqnarray} 
where $V_D$ is the volume of the spacelike sections coming from the space 
integration due to the homogeneity and the isotropy of the FRW metric. 
Assuming $D$ to be an integer, $V_D$ is given by 
\begin{eqnarray}
\label{IV.16}
V_D &=& \cases {\frac{2 \pi^{(\frac{D+1}{2})}}{\Gamma(\frac{D+1}{2})},          
               & if $k=+1$, \cr          
               \frac{\pi^{(\frac{D}{2})}}{\Gamma(\frac{D}{2}+1)}{\chi_c}^D,          
               & if $k=0$, \cr
               \frac{2\pi^{(\frac{D}{2})}}{\Gamma(\frac{D}{2})}f(\chi_c),            
               & if $k=-1$, \cr}
\end{eqnarray}
where $\chi_c$ is a cut--off and $f(\chi_c)$ is a function thereof 
(see Appendix A for more details). Now let us assume $D$ to be a dynamical 
parameter accepting any non--integer value. So, the volume of spacelike 
sections, $V_D$, must be defined for fractal or non--integer dimensions.
Since the volume of a fractal structure having non--integer dimension 
is beyond the scope of this paper, we take the expression (\ref{IV.16}) 
to be valid for the volume of spacelike sections whether the spatial 
dimension is integer or not.\\ 
Now, the action (\ref{IV.15ab}) depends on the second time derivative of the 
dynamical variable $a$. We can get rid of it through introduction of a 
total time derivative, as it was done in section~\ref{subsec-action}. 
We, however, intend to generalize our action to the case of variable space 
dimension. Therefore we have to decide which operation being done 
first: generaliztion to variable space dimension or substitution of the 
second derivative with a total time derivative, as it was done in the case of 
the original decrumpling model in section~\ref{subsec-action} \cite{M3}. 
Substituting first the term $\ddot a$ with a total time derivative we obtain
\begin{eqnarray}
\label{IV.17}
S_I &:=& \frac{1}{2 \kappa}
\int dt \{ \frac{2DV_D}{{a_0}^D}
\frac{d}{dt} (\frac{\dot a a^D}{a N}) -
\frac{V_D D (D-1)}{N} ( (\frac{\dot a }{a})^2 \nonumber\\
&&-\frac{N^2 k }{a^2})
(\frac{a}{a_0})^D-
\rho N V_D (\frac{a}{a_0})^D \}.
\end{eqnarray}
Omitting the total time derivative in the action (\ref{IV.17}), we
reach at the following Lagrangian 
\begin{eqnarray}
\label{IV.18}
L_I &:=& -\frac{1}{2 \kappa} (\frac{a}{a_0})^D \frac{D(D-1)}{N}
((\frac{\dot a}{a})^2-\frac{N^2 k}{a^2})\nonumber\\ 
&&-\rho N V_D (\frac{a}{a_0})^D.
\end{eqnarray}
In the case $k=0$, this is the same Lagrangian as (\ref{III.7}) except for 
introducing the dimensional dependence of the volume and the special form of 
the matter Lagrangian. As we are going to work in the gauge $\dot N = 0$
we could derive first the field equations and then insert the gauge 
condition, or assume first that $N = {\rm constant}$ and then vary the 
Lagrangian. But it is easily seen that both procedures lead to the same 
result. Therefore, we will assume from now on, without loss of generality, 
that $N = {\rm constant}$.\\ 
Now, we assume the space dimension $D$ to be a variable and use the        
constraint (\ref{con}) in order to reduce the dynamical variables           
to $N$ and $a$. The equations of motion are then obtained by variation of the %
Lagrangian (\ref{IV.18}) with respect to $N$ and $a$. For the variation of                      
$\rho$ we do need the following relations based on Eqs.(\ref{IV.4}, 
\ref{IV.5}) and (\ref{IV.6}): 
\begin{eqnarray}                                                            
\label{IVa.18}                                                              
{j^{\alpha}}_{; {\alpha}}=\frac{1}{N V_D (\frac{a}{a_0})^D} \frac{\partial} 
{\partial x^{\alpha}}(N V_D (\frac{a}{a_0})^D j ^{\alpha})=0,\\             
\label{IVb.18}                                                              
\mu ^2= -\frac{1}{N^2 {V_D}^2 (\frac{a}{a_0})^{2D}}                         
[(N V_D (\frac{a}{a_0})^D j ^{\alpha})(N V_D (\frac{a}{a_0})^D j ^{\beta})] 
g_{\alpha \beta},                                                           
\end{eqnarray}                                                               
and                                                                          
\begin{equation}                                                             
\label{IVc.18}                                                               
\delta \mu= -\mu [(\frac{d \ln V_D}{dD}+\ln \frac{a}{a_0})\frac{dD}{da}+     
\frac{D}{a}] \delta a.                                                       
\end{equation}                                                               
As mentioned in (\ref{IVb.12}) variation of $\mu$ with respect to N is   
zero for constant dimension. It is easy to see that this fact is still true
in the case of variable dimension.\\
Now, using (\ref{IVc.18}) and the definition of the energy density and the 
pressure, as given by Eqs.(\ref{IV.8}) and (\ref{IV.9}) the following 
equations of motion are obtained:
\begin{equation}
\label{IV.24}
\frac{1}{N^2}(\frac{\dot a}{a})^2+
\frac{k}{a^2}=\frac{2 \kappa \rho}{D(D-1)},
\end{equation}
and
\begin{eqnarray}
\label{IV.25}
&&(D-1) [\frac{\ddot a}{a}+((\frac{\dot a}{a})^2+\frac{N^2 k}{a^2})
(-\frac{D^2}{2C} \frac{d \ln V_D}{d D}\nonumber\\
&&-1-\frac{D(2D-1)}{2C(D-1)}+\frac{D^2}{2 D_0}) 
+\kappa p N^2 (- \frac{d \ln V_D}{d D} \frac{D}{C}\nonumber\\
&&-\frac{D}{C} \ln{\frac{a}{a_0}}+1)]=0.
\end{eqnarray}
Using Eqs.(\ref{IV.24},\ref{IV.25}), one easily gets the continuity equation 
for our model with variable space dimension:
\begin{equation}
\label{IV.20a}
\frac{d}{dt}(\rho(\frac{a}{a_0})^D V_D)+p \frac{d}{dt}((\frac{a}{a_0})^D V_D)
=0.
\end{equation}
This continuity equation can be integrated for the case of dust or radiation 
to obtain the energy density as a function of time. For radiation era,
$p=\frac{\rho}{D}$, we have
\begin{equation}
\label{IV.21a}
\rho = \rho_0 e^{C(\frac{D}{D_0}-1)}(\frac{D}{D_0})^{\frac{C}{D_0}}                
{\frac{V_{D_0}}{V_D}}e^{-\int^D_{D_0} dD \frac{1}{D}
\frac{d \ln V_D}{d D}},         
\end{equation}
and for matter era, $p=0$:
\begin{equation}
\label{IV.21b}
\rho=\rho_0 e^{C(\frac{D}{D_0}-1)}\frac{V_{D_0}}{V_D}.
\end{equation}
Using Eqs.(\ref{c2}, \ref{IV.24}), the evolution equation for the spatial 
dimension is derived as 
\begin{equation}
\label{IV.26}
{\dot D}^2= \frac{N^2 D^4}{C^2}[\frac{2 \kappa \rho}{D(D-1)}- k {\delta}^{-2}
e^{-2C/D}].
\end{equation}
However, if we assume the time variability of $D$ first, substitution of
the second time derivative by a total time derivative leads to the following 
action
\begin{eqnarray}
\label{IV.19}
S_{II} &:=& \frac{1}{2\kappa} \int dt \{2 \frac{d}{dt}(\frac{DV_D}{N}
\frac{\dot a}{a}(\frac{a}{a_0})^D)-\frac{2 \dot D \dot a V_D}{a N} 
(\frac{a}{a_0})^D\nonumber\\
&&-2\frac{DV_D}{N}\frac{\dot a \dot D}{a} \ln \frac{a}{a_0} 
(\frac{a}{a_0})^D-\frac{2 D \dot D}{N} \frac{\dot a}{a}(\frac{a}{a_0})^D 
\frac{dV_D}{dD}\nonumber\\
&&-\frac{V_DD(D-1)}{N}((\frac{\dot a}{a})^2-\frac{N^2 k}{a^2})
(\frac{a}{a_0})^D\nonumber\\
&&-\rho N V_D (\frac{a}{a_0})^D \}.
\end{eqnarray}
Neglecting now the total time derivative, we reach at the following Lagrangian 
\begin{eqnarray}
\label{IV.20} 
L_{II}&:=&-\frac{V_D}{2\kappa} (\frac{a}{a_0})^D\{\frac{2\dot D \dot a}{a N}+ 
\frac{2 D \dot a \dot D }{aN} \ln \frac{a}{a_0}+\frac{D(D-1)}{N}\nonumber\\
&& \times((\frac{\dot a}{a})^2-\frac{N^2 k}{a^2})+
\frac{2 D \dot D}{N}\frac{\dot a}{a} (\frac{a}{a_0})^D\frac{d \ln V_D}{dD} \}
\nonumber\\
&&-\rho V_D N (\frac {a}{a_0})^D.
\end{eqnarray}
This Lagrangian is more general than $L_{I}$. Putting $\dot D= 0$ in $L_{II}$ 
we obtain the Lagrangian $L_{I}$ which can still be considered as a 
Lagrangian with variable space dimension.\\ 
Using again the dimensional constraint (\ref{con}) and                   
Eqs.(\ref{IV.8} ,\ref{IV.9}) and (\ref{IVc.18}), variation of $L_{II}$   
with respect to $N$ and $a$ leads to the following equations of motion   
\begin{equation}
\label{IV.21}
\frac{1}{N^2}(\frac{\dot a}{a})^2=\frac{2 \kappa \rho-\frac{kD(D-1)}{a^2}}
{D[\frac{2D^2}{D_0}-D-1-\frac{2D}{C}-\frac{2 D^2}{C} \frac{d \ln V_D}{d D}]},
\end{equation}
and
\begin{eqnarray}
\label{IV.22}
&&\frac{\ddot a}{a}[\frac{2D}{C}+\frac{2D^2}{C} \ln{\frac{a}{a_0}}+
\frac{2D^2}{C}\frac{d \ln V_D}{d D}-D+1] \nonumber\\
&&+(\frac{\dot a}{a})^2[-\frac{D^4}{C^2}(\ln \frac {a}{a_0})^2+
(-\frac{4D^3}{C^2}+\frac{3D^3}{2C}-\frac{5D^2}{2C} \ln \frac{a}{a_0}
\nonumber\\
&&-\frac{d \ln V_D}{d D}( \frac{4D^3}{C^2}-\frac{3D^3}{2C}+\frac{5D^2}{2C}+
\frac{2D^4}{C^2} \ln{\frac{a}{a_0}})\nonumber\\
&&-\frac{D^4}{C^2 V_D} \frac{d^2 V_D}{d D^2}+\frac{3D^2}{C}- \frac{2D^2}{C^2}
-\frac{5D}{2C}-\frac{(D-1)(D-2)}{2}]\nonumber\\
&&+\frac{k N^2}{a^2}[\frac{D(2D-1)}{2C}+\frac{D^2(D-1)}{2C}
\frac{d \ln V_D}{d D}\nonumber\\
&&-\frac{(D-1)(D-2)}{2} +\frac{D^2(D-1)}{2C}\ln \frac{a}{a_0}]+
N^2 \kappa p (\frac{D}{C} \ln \frac{a}{a_0}\nonumber\\
&&+\frac{D}{C}\frac{d \ln V_D}{d D}-1)=0.
\end{eqnarray}
In the limit $ C \rightarrow +\infty$ and $D=D_0={\rm const}$, Eqs.(\ref{IV.24},
\ref{IV.25}, \ref{IV.21}, \ref{IV.22}) reduce to the standard Friedmann 
equations (\ref{III.8},\ref{III.10}). Using Eqs.(\ref{IV.21}, \ref{IV.22}), 
one easily gets the continuity equation for our model with variable space 
dimension. Now, using Eqs.(\ref{c2}, \ref{IV.21}), the evolution equation 
for the spatial dimension is derived as 
\begin{equation}
\label{IV.23}
{\dot D}^2=\frac{N^2 D^3[2 \kappa \rho - k D(D-1) {\delta}^{-2} e^{-\frac{2C}
{D}}]}
{C^2[\frac{2D^2}{D_0}-D-1-\frac{2D}{C}-\frac{2 D^2}{C} \frac{d \ln V_D}{d D}]}.
\end{equation}
From Eqs.(\ref{IV.26}) and (\ref{IV.23}), it is easy to see that in the limit 
$C \rightarrow +\infty$, we have $\dot D=0$, which corresponds to the standard
universe model with constant dimension.
To understand the dynamical behavior of dimension in the models discussed 
we have to substitute $\rho$ and $V_D$ from (\ref{IV.21a}--\ref{IV.21b}) 
and (\ref{IV.16}) in (\ref{IV.26}) or (\ref{IV.23}) respectively. 
In the case of $k = 0$ it is, however, seen easily that
there is no turning point in dimension and both models can be contracted
up to $a = \delta$ corresponding to $D \rightarrow +\infty$. The discussion
can be simplified for a de Sitter--like model which we will discuss in the 
next section.

\subsection{de Sitter--like model}
\label{subsec-desitter}

Consider a universe dominated by the constant vacuum energy, $\rho_{\Lambda}$, 
corresponding to a cosmological constant $\Lambda$ and the equation of state
$p_{\Lambda} = - \rho_{\Lambda}$. Note that these relations are satisfied 
by the continuity equation (\ref{IV.20a}). Substituting 
\begin{equation}
\label{IV.27}
\rho_{\Lambda} \equiv \frac{\Lambda}{8 \pi G},
\end{equation}
in (\ref{IV.26}) and (\ref{IV.23}), we obtain evolution equation of the 
spatial dimension for the Lagrangian $L_{I}$ 
\begin{equation}
\label{IV.28}
{\dot D}^2= \frac{-N^2 D^3}{C^2 (D-1)}[D(D-1){\delta}^{-2}e^{-2C/D}-
2 \Lambda] \equiv -V_{I}(D),
\end{equation}
and for the Lagrangian $L_{II}$
\begin{equation}
\label{IV.29}
{\dot D}^2=-\frac{N^2 D^3 [D(D-1) \delta^{-2} e^{-2C/D} -2 \Lambda]}
{C^2 [\frac{2D^2}{D_0}-\frac{2D^2}{C} \frac {d \ln V_D}{dD}- \frac{2D}{C}
-D-1]} \equiv - V_{II}(D),
\end{equation}
where we have set $k = +1$. As was mentioned in Sec.~\ref{subsec-field},
these equations may be interpreted as ``energy equation'' 
for $D$, having vanishing total energy with the potentials 
$V_I(D)$ and $V_{II}(D)$ respectively. Both models have a ``classical'' 
turning point at $D_T$, where $\dot D$ vanishes. This is the point of maximum
dimension of space corresponding to the minimum radius of the universe
$a_T$ (see Fig. 5). We obtain this dimension  as the solution of the equation 
\begin{equation}
\label{IV.30}
D(D-1){\delta}^{-2} e^{-2C/D} -2 \Lambda=0.
\end{equation}
In what follows we will choose for the cosmological constant  
\begin{equation}
\label{IV.30a}
\Lambda=3 G^{-1}=3 {l_{Pl}}^{-2}.
\end{equation}
Eq.(\ref{IV.30}) has to be solved numerically. 
Table III shows some interesting $D_T$ values.
There is a minimum dimension for both models which is reached asymptotically.
This minimum dimension for model I is $D = 1$ corresponding to a maximum 
radius $a_{Im}\equiv \delta e^C$. To study the asymptotic behavior of 
$V_{II}(D)$, we need to calculate the term 
$\frac {d \ln V_D}{d D}$. From (\ref{IV.16}), for $k=+1$, we obtain 
\begin{equation}
\label{IV.31}
\frac{d \ln V_D}{d D}=\frac{1}{2}( \ln \pi -\psi (\frac{D+1}{2})),
\end{equation}
where $\psi$ is the logarithmic derivative of the gamma function, $\psi 
\equiv \frac{\Gamma '(x)}{\Gamma(x)}$, called Euler's Psi function.
Substituting (\ref{IV.31}) into (\ref{IV.29}), it is easy to see that the 
potential $V_{II}(D)$ has an asymptotic behavior at $D_{IIm}$, corresponding
to $a_{IIm}$, being the solution of the equation
\begin{equation}
\label{IV.32}
\frac{2 D^2}{D_0} -\frac{D^2}{C} (\ln \pi -\psi(\frac{D+1}{2})) 
-\frac{2 D}{C}-D-1=0.
\end{equation}
Some numerical values for $D_{IIm}$ and $a_{IIm}$ are also given in 
Table III.
A generic example of the potential $V_{II}(D)$ is plotted in Fig. 5. It 
exhibits the classically allowed region (${\dot D}^2 >0$). Therefore in a 
de Sitter--like universe, taking the model II, for the dimension of space  
we have $D_{IIm}< D<D_T$. $D_T$, which corresponds to the minimum contraction  
in the de Sitter--like model, can have very different dimensions, as 
given in Table III. There we can also see that the minimum scale factor,
$a_T$, corresponding to $D_T$ can be up to about 10 times the Planck 
length, assuming $D_0 = 3$. The assumption $D_0 = 2$ leads to almost the 
same result with values for $D_T$ a bit smaller than those of Table III.
The maximum radius $a_{IIm}$, corresponding to the minimum dimension
$D_{IIm}$ which is about 2 as can be seen from Table III, is many order 
of magnitudes bigger than the present radius of the universe.\\ 
The behavior of $V_I(D)$ is qualitatively the same as $V_{II}(D)$,
except that the minimum dimension is at $D_{IIm} \equiv 1$. In the next 
sections on quantum cosmology of our de Sitter--like models we will 
come back to the details of this behavior. 
The detailed behavior of the potentials and the field equations leads
to interesting cosmological consequences which we will discuss in a 
forthcoming paper \cite{rman}.\\
\small{
TABLE III. $a_T/l_{Pl}, D_T, \log(a_{IIm}/l_{Pl}), D_{IIm}, \log(a_{Im}/l_{Pl})$
for $k=+1$ and for interesting values of $C$ corresponding to $D_{Pl}=3.001, 4, 10,
25$ and $D_0 \simeq 3$ (cf Table I for corresponding values 
of $\delta$).
}
\begin{center}  
\begin{tabular}{|p{2.5 cm}|p{1.5 cm}|p{1.5 cm}|p{2.5 cm}|p{1.5 cm}|p{2.5 cm}|}  \hline\hline
$C$ & $a_T/l_{Pl}$ & $D_T$ & $\log(a_{IIm}/l_{Pl})$ & $D_{IIm}$ & $\log(a_{Im}/l_{Pl})$  \\ \hline
 $1.2588 \times 10^6$ & $1.0$ &  $3.001$ & $7.8160\times 10^4$ & $2.0$ & $3.6452\times10^5$ \\  \hline
 $1.6788 \times 10^3$ &  $1.4$ &  $3.996$ &  $1.5079\times10^2$ & $2.189$ & $5.4682\times 10^2$ \\  \hline
 $5.9957 \times 10^2$ &  $3.0$ &  $9.782$ & $9.2644\times10$ & $2.194$ & $2.3435\times 10^2 $\\  \hline
 $4.7693 \times 10^2$ &  $8.0$ &  $22.42$ & $8.6034\times10$ & $2.196$ & $1.9884\times 10^2$ \\ \hline \hline
\end{tabular}
\end{center}

\section{WAVE FUNCTION OF OUR MODEL UNIVERSE} \label{sec-wave}

We are now interested in quantum cosmological behavior of our model universe
and its differences to the Wheeler--DeWitt equation and its 
solutions for the de Sitter minisuperspace in 3--space. 
We will therefore use the canonical approach of quantization due 
to DeWitt and Wheeler \cite{bsde,cwmi}. To do so, let us first briefly 
review the canonical quantization of the universe with constant spatial 
dimension. 

\subsection{Brief review of the tunneling and the Hartle-Hawking wave 
function} \label{subsec-brief}

According to the quantum approach to the standard cosmology 
\cite{{avil},{jbha},{adli}}, a small closed universe can spontaneously
nucleate out of ``nothing'', where by ``nothing'' we mean a state with no 
classical space--time. The cosmological wave function can be used to 
calculate the probability distribution for the initial configuration of the 
nucleating universe. Once the universe nucleated, it is expected to go 
through a period of inflation.\\
Let us now illustrate how the nucleation of the universe can be described 
in the simplest de Sitter minisuperspace model. In this model, the  
universe is assumed to be homogeneous, isotropic, closed, and filled with a 
vacuum of constant energy density $\rho_{\Lambda}$. The universe should be  
closed, since otherwise its volume will be infinite and the nucleation 
probability would be zero. The radius of the 
universe $a$ is the only dynamical variable of the model, and the wave 
function $\Psi(a)$ satisfies the Wheeler--DeWitt (WDW) equation:
\begin{equation}
\label{V.1}
H \Psi(a) = 0,
\end{equation}
where $H$ is the corresponding Hamiltonian of the model. In our de Sitter
case it is written as
\begin{equation}
\label{V.2}
[\frac{d^2}{d a^2}-\frac{9 \pi^2 a^2}{4 G^2}(1-l_{\Lambda}^{-2} a^2)]
\Psi(a)=0.
\end{equation}
Here 
\begin{equation}
\label{V.2a}
l_{\Lambda}^{-2} \equiv \frac{8 \pi G \rho_{\Lambda}}{3}=\frac{\Lambda}{3}.
\end{equation}
The value $\Lambda=3G^{-1}$ corresponds to $l_{\Lambda}=l_{Pl}$.
We have disregarded the ambiguity in the ordering of non--commuting 
operators $a$ and $d/da$, which is unimportant in the semiclassical 
approximation we are using here. Eq.(\ref{V.2}) has the form of 
a one--dimensional Schr\"{o}dinger equation for a ``particle'' described 
by a coordinate $a(t)$ having zero energy and moving in the potential 
\begin{equation}
\label{V.2b}
U(a)=\frac{9 \pi^2 a^2}{4 G^2}(1-l_{\Lambda}^{-2} a^2).
\end{equation}
The classically allowed region is $a \geq l_{\Lambda}$, and the WKB solution 
of Eq.(\ref{V.2}) in this region is 
\begin{equation}
\label{V.3}
\Psi_{\pm}(a)=[p(a)]^{-1/2}\exp[\pm i \int^a_{l_{\Lambda}}p(a') da'\mp
i\pi /4],
\end{equation}
where $p(a)=[-U(a)]^{1/2}$. The under--barrier ($a < l_{\Lambda}$) solutions 
are 
\begin{equation}
\label{V.4}
\tilde{\Psi}_{\pm}(a)= |p(a)|^{-1/2} \exp[ \pm \int^{l_{\Lambda}}_a |p(a')|da'].
\end{equation}
The classical momentum conjugate to $a$ is $p_a=-\frac{3 \pi a \dot a}{2 G N}$. For 
$a \gg l_{\Lambda}$ we have 
\begin{equation}
\label{V.5}
{\hat {p}}_a \Psi_{\pm}(a) \approx \pm p(a) \Psi_{\pm}(a),
\end{equation}
where $\hat{p}_a= -i \partial /\partial a$.
Thus $\Psi_{-}(a)$ and $\Psi_{+}(a)$ describe an expanding and a 
contracting universe, respectively. According to Vilenkin, the tunneling 
boundary condition 
\cite{avile} requires that only the expanding component should be present 
at large $a$. Therefore we obtain for the Vilenkin tunneling wave 
function $\Psi_{V}$
\begin{equation}
\label{V.6}
\Psi_{V}(a > l_{\Lambda})=\Psi_{-}(a).
\end{equation}
The under-barrier wave function is then found from WKB connection formula:
\begin{equation}
\label{V.7}
\Psi_{V}(a<l_{\Lambda}) =\tilde {\Psi}_{+}(a)-\frac{i}{2} \tilde{\Psi}_{-}(a).
\end{equation}
The growing exponential $\tilde {\Psi}_{-}(a)$ and the decreasing exponential 
$\tilde \Psi_{+}(a)$ have comparable amplitudes at the nucleation point 
$a = l_{\Lambda}$, but away from that point the decreasing exponential 
dominates (see Fig. 6). Therefore, the nucleation probability can be 
approximated as \cite{17}
\begin{equation}
\label{V.8}
{\cal{P}} \sim |\frac{\Psi_V(l_{\Lambda})}{\Psi_V(0)}|^2 \sim 
\exp [-2 \int^{l_{\Lambda}}_0 
|p(a')| da'] = \exp(-\frac{3 \pi}{G \Lambda}). 
\end{equation}
The Hartle--Hawking wave function satisfying the no--boundary 
boundary condition takes the from \cite{jbha}
\begin{equation}
\label{V.13}
\Psi_{HH}(a<l_{\Lambda})=\tilde \Psi_{-}(a),
\end{equation}
for the under--barrier wave function and
\begin{equation}
\label{V.14}
\Psi_{HH}(a>l_{\Lambda})=\Psi_{+}(a)+\Psi_{-}(a),
\end{equation}
in the classically allowed range. This wave function describes a contracting 
and re--expanding universe; under the barrier $\Psi_{HH}(a)$ is exponentially
suppressed (see Fig. 6).\\ 
Using the anti--Wick rotation for Euclideanization of a Lorentzian path 
integral, Linde \cite{adli} suggested his tunneling wave function.
Linde's wave function has only the decreasing exponential 
${\tilde {\Psi}}_{+}(a)$ in the non--classically allowed region:
\begin{equation}
\label{lin.1}
\Psi_{L}(a <l_{\Lambda})={\tilde{\Psi}}_{+}(a).
\end{equation}
The continuation to the classically allowed region gives
\begin{equation}
\label{lin.2}
\Psi_{L}(a >l_{\Lambda})=\frac{1}{2}[\Psi_{+}(a)+\Psi_{-}(a)].
\end{equation}
Different probability densities are assumed to correspond to different 
boundary conditions implied by the Hartle--Hawking ``no--boundary'' 
proposal \cite{jbha}, the ``tunneling'' proposal of Vilenkin
\cite{{avile}, {30}}, and that of Linde \cite{adli}. In particular,
the nucleation probability for instanton--dominated transitions is assumed 
to be  
\begin{eqnarray}
\label{lin.3}
{\cal{P}} \propto {|\Psi|}^2 \propto
\cases { \exp(-2S_E) \, & for $\Psi_{HH}$,  \cr       
         \exp(2S_E)     \, & for $\Psi_L,\,\Psi_{V}$, \cr}
\end{eqnarray}
where $S_E$ is the Euclidean action of the instanton. 
Problems concerning the interpratation of the  
wave function of the universe are studied in many articles \cite{rpar}. 
For a recent discussion of problems associated with defining the initial 
cosmological wave function see e.g. Ref.\cite{33}. The debate about 
the form of the wave function of the universe has recently intensified by 
Vilenkin \cite{34}, Hawking and Turok \cite{35}, and Linde \cite{36};
see also Refs.\cite{{37}, {38}}.

\subsection{Wheeler--DeWitt equation in the universe with dynamical 
spatial dimension}

Turning now to the canonical quantization of our model with variable 
space dimension, we consider the simplest case of the de Sitter 
minisuperspace. We therefore take $k=+1$ and 
$\rho=\rho_{\Lambda}=\frac {\Lambda}{8 \pi G}$, in the Lagrangian $L_I$ 
and $L_{II}$. The corresponding Hamiltonians to $L_I$ and $L_{II}$ can be 
written in the form 
\begin{eqnarray}
\label{V.10}
H_{I}&:=& N {\cal {H}}_{I} \nonumber\\
&:=&p_{Ia} \dot a - L_I \nonumber\\
&=&N \{ - \frac{\kappa a^2 
{p_{Ia}}^2}{2 V_D (\frac{a}{a_0})^D D (D-1)}
-\frac{ V_D D (D-1)}{2 \kappa a^2} (\frac{a}{a_0})^D \nonumber\\
&+& \frac{V_D \Lambda}{\kappa} (\frac{a}{a_0})^D \}, 
\end{eqnarray}
and
\begin{eqnarray}
\label{V.10a}
H_{II}&:=&N {\cal {H}}_{II} :=p_{IIa} \dot a - L_{II} \nonumber\\
&=&N \{ - \frac{\kappa a^2 {p_{IIa}}^2}{4 D V_D (\frac{a}{a_0})^D 
[\frac{D^2}{D_0}
-\frac{D^2}{C} \frac{ d \ln V_D}{d D}-\frac{D}{C}
-\frac{D+1}{2}]}\nonumber\\
&-&\frac{D(D-1)V_D}{2 \kappa a^2} (\frac{a}{a_0})^D 
+\frac{V_D \Lambda}{\kappa}
(\frac{a}{a_0})^D \},
\end{eqnarray}
where $V_D$ is the volume of D--sphere $S^D$ (cf. Appendix A and 
Eq.\ref{IV.16}). The canonical momenta $p_{Ia}$ and $p_{IIa}$ are defined as
\begin{eqnarray}
\label{V.12}
p_{Ia}&:=&\frac{\partial L_I}{\partial \dot a} \nonumber\\
&=&- \frac{V_D D (D-1) \dot a}{\kappa N a^2}(\frac{a}{a_0})^D,
\end{eqnarray}
and 
\begin{eqnarray}
\label{V.13a}
p_{IIa}&:=&\frac{\partial L_{II}}{\partial \dot a}+\frac
{\partial L_{II}}{\partial \dot D} 
\frac{\partial \dot D}{\partial \dot a}\nonumber\\
&=&-\frac{V_D D \dot a}{\kappa N a^2} (\frac{a}{a_0})^D 
[\frac{2D^2}{D_0}-\frac{2D^2}{C} \frac{d \ln V_D}{d D}\nonumber\\
&-&\frac{2D}{C}-D-1].
\end{eqnarray}
The above systems can be quantized by the assignment
\begin{equation}
\label{p}
{p_{Ia}}^2,\;{p_{IIa}}^2 \to -{{a_0}^{-2D_0}} \frac{1}{a^q}
\frac{\partial}{\partial a}a^q \frac{\partial}{\partial a},
\end{equation}
where $q$ is the factor ordering parameter. Note that we have introduced the 
coefficient ${a_0}^{-2D_0}$ in (\ref{p}) for the kinetic term in 
WDW equation to have the right dimension. The WDW equation is obtained 
by applying the classical constraint 
\begin{eqnarray}
\label{V.15}
\frac { \delta H_I}{\delta N}=0,\\
\frac { \delta H_{II}}{\delta N}=0,
\end{eqnarray}
on the wave function $\Psi_I(a)$ and $\Psi_{II}(a)$ respectively:
\begin{eqnarray}
\label{V.16}
{\cal{H}}_{I} \Psi_{I} (a)&=&\{ \frac{1}{a^q} \frac{\partial}{\partial a}a^q 
\frac{\partial}{\partial a}-\frac{{a_0}^{2D_0} {V_D}^2 D(D-1)}{\kappa ^2 a^2}
(\frac{a}{a_0})^{2D}\nonumber\\
&\times&(\frac{D(D-1)}{a^2}-2 \Lambda) \}\Psi_I(a)=0,  
\end{eqnarray}
and
\begin{eqnarray}
\label{V.16a}
{\cal{H}}_{II} \Psi_{II} (a)&=&\{ \frac{1}{a^q} \frac{\partial}{\partial a}a^q 
\frac{\partial}{\partial a}
-\frac{{a_0}^{2D_0} {V_D}^2 D}{\kappa ^2 a^2}(\frac{a}{a_0})^{2D}\nonumber\\
&\times&(\frac{D(D-1)}{a^2}-2 \Lambda) [\frac{2D^2}{D_0}
-\frac {2 D^2}{C} \frac{d \ln V_D}{d D}\nonumber\\
&-& \frac{2D}{C}-D-1]  \} \Psi_{II}(a)=0. 
\end{eqnarray}
In the semiclassical approximation the ambiguity in the ordering of $a$ 
and $d/da$ can be ignored. Therefore, we will assume $q=0$. The corresponding potentials for 
$\Psi_I(a)$ and $\Psi_{II}(a)$ are  
\begin{eqnarray}
\label{V.17}
U_{I}(a)&:=& \frac{{a_0}^{2D_0} {V_D}^2 D (D-1)}{\kappa ^2 a^2}
(\frac{a}{a_0})^{2D}\nonumber\\
&\times&(\frac{D(D-1)}{a^2}-2 \Lambda),\\
U_{II}(a)&:=&\frac{{a_0}^{2D_0} {V_D}^2 D}{\kappa ^2 a^2}
(\frac{a}{a_0})^{2D}(\frac{D(D-1)}{a^2}-2 \Lambda)
[\frac{2D^2}{D_0}\nonumber\\
&-&\frac{2 D^2}{C} \frac{d \ln V_D}{d D}- \frac{2D}{C}-D-1].
\end{eqnarray}
It is worth noticing that in these equations $D$ is a function of 
the scale factor, $a$, according to the constraint (\ref{con}).
It can be easily shown 
$$U_I(a) \to U(a),\;\;\;\;\;\;\;{\rm as}\;\;D=D_0=3,$$
and 
$$U_{II}(a) \to U(a),\;\;\;\;\;\;\;{\rm as}\;\;D=D_0=3,\;\;C \to \infty.$$
Note that $C$ appears explicitly in $U_{II}(a)$. This is due to appearance of
the time derivative of $D$ in the Lagrangian $L_{II}$. The shape of the 
potential $U_I(a)$ and $U_{II}(a)$ determine how we impose the appropriate 
boundary condition and evaluate the wave function of our model. Using 
(\ref{IV.16}) for $k=+1$ and (\ref{IV.31}) and the dimensional constraint 
(\ref{con}), the potentials $U_I(a)$ and $U_{II}(a)$ can be written in terms 
of $D$ only:
\begin{eqnarray}
\label{V.18}
U_I(D)&=&\frac{4 {a_0}^{2D_0} \pi^{(D+1)} D(D-1) \delta^{-2} 
e^{2C(1-\frac{D}{D_0}
-\frac{1}{D})}}{\kappa ^2 (\Gamma(\frac{D+1}{2}))^2}\nonumber\\
&\times& (D(D-1) \delta^{-2}e^{-2C/D}-2 \Lambda),
\end{eqnarray}
and
\begin{eqnarray}
\label{V.19}
U_{II}(D)&=&\frac{4 {a_0}^{2D_0} \pi^{(D+1)} D \delta^{-2}
e^{2C(1-\frac{D}{D_0}-\frac{1}{D})}}{ \kappa^2 (\Gamma(\frac{D+1}{2}))^2}\nonumber\\
&\times&(D(D-1) \delta^{-2} e^{-2C/D}- 2 \Lambda) ( \frac {2D^2} {D_0}\nonumber\\
&-&\frac{D^2}{C}( \ln \pi -\psi(\frac{D+1}{2}))-\frac{2D}{C}-D-1).
\end{eqnarray}
Let us now obtain the zero points of $U_I(D)$ and $U_{II}(D)$. Using the 
following asymptotic expression for $\Gamma$ and $\psi$ function as 
$D \to +\infty$ 
$$\Gamma(D) \to e^{-D} D^{D-\frac{1}{2}} \sqrt {2 \pi}(1+\frac{1}{12D}+...),$$
and
$$\psi(D) \to \ln D -\frac{1}{2D} - \sum_{k=0}^{\infty} \frac{B_{2k}}
{2k D^{2k}},$$
where $B_{2k}$ are the Bernoulli numbers, we obtain for 
$D \to +\infty$ 
$$U_I(D) \rightarrow  {\frac {2 {a_0}^{2 D_0} }{\kappa ^2 \delta^4} (2 \pi)^D D^{(4-D)}
e^{ D ( 1- \frac {2C} {D_0} ) }} \to 0, $$
and 
$$
U_{II}(D) \rightarrow {\frac{2 {a_0}^{2 D_0}}{\kappa ^2 \delta^4 C}(2 \pi)^D
D^{(6-D)} e^ {D(1-\frac{2C}{D_0})} \ln (\frac{D}{2}) } \rightarrow 0.
$$
Therefore, the potentials $U_I(D)$ and $U_{II}(D)$ both tend to zero as 
$D \to +\infty$. According to Eqs.(\ref{IV.28}--\ref{IV.32}) and using 
Eqs.(\ref{V.18},\ref{V.19}) three other zero points of $U_I(D)$ 
and $U_{II}(D)$ are obtained. Those of $U_I(D)$ are
\begin{eqnarray}
\label{V.20}
D&=& \cases { D\rightarrow +\infty\, & for $a\rightarrow \delta$,  \cr       
              D_T\, & for $a=a_T$,  \cr 
              1\, & for $a=a_{Im}$, \cr
              0\, & for $a \to +\infty$. \cr}
\end{eqnarray}
For $U_{II}(D)$ we have the same zeros except the third one which is at 
$D=D_{IIm}$ corresponding to $a=a_{IIm}$, whose corresponding values are
given in Table III.\\
We are now in a position to compare these potentials with the potential
of the standard de Sitter minisuperspace depicted in Fig. 6.
To make the comparison as simple as possible, we draw
the dependence of our potential $U_I(a)$ against the scale factor $a$. 
As we see from Fig. 7, the behavior of the potential $U_I(a)$ for 
small scale factors of the order of the Planck length is similar to the
potential $U(a)$. Instead, for large scale factors we see a completely 
different behavior. Here we have two potential barriers with different
heights depending on the value of $C$. For example, for $C=1678$, 
corresponding to $D_{Pl} = 4$, the ratio of their heights is of the order of
$10^{248}$.\\
The height of the potential barrier in the region $\delta<a<a_T$ is also 
interesting to be compared to that of the potential $U(a)$. From 
Eq.(\ref{V.2b}), it is easy to see that the height of the potential 
barrier of $U(a)$, in the region $0<a<l_{\Lambda}$, is of the order $G^{-1}$,
or the square of the Planck energy, which is of the order of $10^{39} GeV^2$. 
In contrast, the height of the potential barrier of $U_I(a)$ in the region 
$\delta <a< a_T$ is about $10^{-158} GeV^2$ for $C= 1678$ and 
$10^{-2361}GeV^2$ for $C=477$. Therefore as $C$ increases, the height of 
the barrier increases so that for $C \simeq 10^6$ it is of the order $G^{-1}$ 
comparable to the potential barrier $U(a)$. The behavior of $U_{II}(a)$ is 
similar to $U_I(a)$. 

\subsection{Wave function of the universe with dynamical space
dimension} \label{subsec-wave}

Here, we are interested in the solutions of the WDW equation in our model 
universe. Let us first use the semiclassical approximation to solve Eqs.
(\ref{V.16}) and then Eq.(\ref{V.16a}). The classically allowed region is $a_T<a<a_{Im}$, 
and the WKB solutions are
\begin{equation}
\label{IV.60}
\Psi_{I \pm}(a)=[p_I(a)]^{-1/2} \exp[\pm i \int_{a_T}^a p_I(a')da' \mp 
i\pi /4],
\end{equation}
where $p_I(a)=[-U_I(a)]^{1/2}$. Under the potential barrier in the region 
$\delta <a <a_T$ the WKB solutions are 
\begin{equation}
\label{IV.61}
\tilde \Psi _{I \pm }(a)= |p_I(a)|^{-1/2} \exp [\pm \int_a ^{a_T} 
|p_I(a')| da'].
\end{equation}
Finally, the solutions for the other under--barrier region $a> a_{Im}$ are
\begin{equation}
\label{IV.62}
\hat \Psi _{I \pm }(a)= |p_I(a)|^{-1/2} \exp [\pm \int_{a_{Im}} ^a 
|p_I(a')| da'].
\end{equation}
From Eq.(\ref{V.12}) we see that the classical momentum 
conjugate to $a$, $p_{Ia}$, is proportional to $- \dot a (D-1)$. For $a_T 
\ll a < a_{Im}$, Eq.(\ref{V.5}) can be used for $\Psi_{I \pm}$:
\begin{equation}
\label{IV.63}
{\hat {p}}_a \Psi_{I \pm}(a) \approx \pm p_I(a) \Psi_{I \pm}(a),
\end{equation}
Thus $\Psi_{I-}(a)$ and $\Psi_{I+}(a)$ describe an expanding and a 
contracting universe, respectively. 
Now, we have to impose boundary condition in order to specify 
the wave function of our model uniquely. We choose, in accordance with the 
tunneling and no--boundary boundary condition, the following boundary 
condition as the scale factor tends to infinity:
\begin{equation}
\label{IV.63c}
\lim_{a \to +\infty} \Psi_I(a) \to 0.
\end{equation}
Another boundary condition must be imposed at the turning point $a=\delta$
where the semiclassical approximation is bound to be wrong. Taking into 
account the normalization of the wave function \cite{sbis}, 
we can write it as 
\begin{equation}
\label{IV.63d}
\Psi _I(\delta) = 1.
\end{equation}
This satisfies the regularity of the wave function, i.e.$|\Psi| <\infty$,
as in the case of Hartle--Hawking \cite{jbha} and Vilenkin \cite{30}
boundary conditions.
According to (\ref{IV.63c}) the growing exponential $\hat \Psi_{I+}(a)$ should 
be absent under the barrier in the region $a>a_{Im}$ : 
\begin{equation}
\label{IV.65}
\Psi_{I}(a > a_{Im})= \hat \Psi_{I-}(a).
\end{equation}
In the classically allowed region, the wave function is found by using the WKB 
connection formula at the turning point $a=a_{Im}$ \cite{emer}:    
\begin{eqnarray}
\label{IV.66}
&&\Psi_{I}(a_T<a<a_{Im})\nonumber\\
&=&2[p_I(a)]^{-1/2} \cos (\int_a^{a_{Im}}p_I(a')da'
-\frac{\pi}{4}).
\end{eqnarray}
This may also be written as
\begin{eqnarray}
\label{IV.67}
&&\Psi_{I}(a_T<a<a_{Im})=2[p_I(a)]^{-1/2}\nonumber\\ 
&\times& \sin (\int_{a_T}^{a_{Im}} p_I(a')da'
-\int_{a_T}^a p_I(a')da'+\frac{\pi}{4})\nonumber\\
&=&2[p_I(a)]^{-1/2} \{ \sin(\int_{a_T}^{a_{Im}}p_I(a')da') \nonumber\\
&\times&\cos(\int_{a_T}^a p_I(a')da'-\frac{\pi}{4})
-\cos(\int_{a_T}^{a_{Im}}p_I(a')da') \nonumber\\
&\times& \sin(\int_{a_T}^a p_I(a')da'-\frac{\pi}{4}) \}.
\end{eqnarray}
Using (\ref{IV.60}), this gives 
\begin{eqnarray}
\label{IV.68}
&&\Psi_{I}(a_T< a<a_{Im})=
\Psi_{I-}(a) e^{i(\int_{a_T}^{a_{Im}} p_I(a')da'
-\frac{\pi}{2})}\nonumber\\
&+&\Psi_{I+}(a)e^{-i(\int_{a_T}^{a_{Im}} p_I(a')da'
-\frac{\pi}{2})}. 
\end{eqnarray} 
Note that the dynamical character of the space dimension makes the 
wave function in the classical region a mixture of expanding and contracting
part. It is in contrast to the standard case of constant space dimension where 
we can choose the wave function to be just expanding as assumed in
the Vilenkin's tunneling boundary condition \cite{avile}. 
The presence of the expanding and contracting component
in the classical region is due to the special form of our potential 
(see Fig. 7). 
The under--barrier wave function is found by using the WKB connection 
formula at the turning point $a=a_T$:
\begin{eqnarray}
\label{IV.69}
&&\Psi_{I}(\delta <a<a_T)=|p_I(a)|^{-1/2} \{ \sin(\int_{a_T}^{a_{Im}}
p_I(a')da') \nonumber\\ 
&\times& \exp (-\int_a^{a_T} |p_I(a')| da')+ \cos (\int_{a_T}^{a_{Im}}
p_I(a') da')\nonumber\\ 
&\times& \exp(\int_a^{a_T} |p_I(a')|da') \}.
\end{eqnarray}
From (\ref{IV.61}), this may be written as
\begin{eqnarray}
\label{IV.70}
\Psi_{I}(\delta <a<a_T)&=&\sin (\int_{a_T}^{a_{Im}}p_I(a') da') \tilde 
\Psi _{I-}(a)\nonumber\\
&+& \cos (\int_{a_T}^{a_{Im}}p_I(a') da') \tilde \Psi_{I+}(a).
\end{eqnarray}
The wave function $\Psi_I(a)$ is schematically represented in Fig. 7.
Similarly, we obtain the following wave functions for the potential 
$U_{II}(a)$: 
\begin{eqnarray}
\label{IV.73}
&&\Psi_{II}(\delta<a<a_T)=\sin (\int_{a_T}^{a_{IIm}}p_{II}(a') da')
\tilde \Psi _{II-}(a)\nonumber\\
&+& \cos (\int_{a_T}^{a_{IIm}}p_{II}(a') da') 
\tilde \Psi_{II+}(a),\\
\label{IV.73a}
&&\Psi_{II}(a_T< a<a_{IIm})= \Psi_{II-}(a) e^{i(\int_{a_T}^{a_{IIm}} 
p_{II}(a')da'-\frac{\pi}{2})}\nonumber\\
&+&\Psi_{II+}(a)e^{-i(\int_{a_T}^{a_{IIm}} 
p_{II}(a')da'-\frac{\pi}{2})},
\end{eqnarray}
and
\begin{equation}
\label{IV.73b}
\Psi_{II}(a > a_{IIm})= \hat \Psi_{II-}(a).
\end{equation}
The above wave functions are different from those of Vilenkin, 
Hartle--Hawking, or Linde. But, as we will see, there are similarities 
between our general wave functions given by 
(\ref{IV.65}, \ref{IV.68}, \ref{IV.70}) for $U_I$ or 
(\ref{IV.73}, \ref{IV.73a}, \ref{IV.73b}) for $U_{II}$ and those of Vilenkin, 
Hartle--Hawking, and Linde. We notice first that our wave functions are real 
for every dimension, even for the limiting case of $C \to +\infty$
or $D \to {D_0}$. Therefore, we conclude that our general wave functions 
are not of Vilenkin's type which is complex. In the classically allowed region, 
$\Psi_{HH}$ and $\Psi_L$ are both a superposition of the expanding and
contracting terms with equal coefficients. In contrast, in our general 
wave functions the coefficients of expanding and contracting components
are different. Looking at our general wave functions, wee see that both 
decreasing and increasing exponential terms are present in the 
non--classical range of $\delta <a <a_T$.
These means that in general, in this region, we expect our wave
functions are a superposition of $\Psi_{HH}$ and $\Psi_{L}$. Requiring the
decreasing term, ${\tilde{\Psi}}_{I+}$, in (\ref{IV.70}) to vanish we are led
to:
\begin{equation}
\label{IV.74}
\int_{a_T}^{a_{Im}} p_I(a')da'=(2n+\frac{1}{2})\pi,
\end{equation}
where $n$ is an integer number. This requirement makes the coefficients of
the expanding and contracting part of our wave functions in the classical
region to be equal. Hence, assuming the relation (\ref{IV.74}) our general
wave functions behave as $\Psi_{HH}$ in the region $\delta <a <a_{Im}$.\\
Now, if we require the increasing term ${\tilde{\Psi}}_{I-}$ to vanish 
under the potential barrier in the region of $\delta <a <a_T$, we obtain
\begin{equation}
\label{IV.74b}
\int_{a_T}^{a_{Im}} p_I(a')da'=2 m \pi,
\end{equation}
where $m$ is an integer number. This leads to the following wave functions:
\begin{eqnarray}
\label{IV.75}
\Psi_{I}^{(m)}(\delta <a<a_T) \equiv \tilde \Psi_{I+}(a),\\
\label{IV.76}
\Psi_{I}^{(m)}(a_T<a<a_{Im}) \equiv i\,[\Psi_{I+}(a)-\Psi_{I-}(a)],\\
\label{IV.76a}
\Psi_{I}^{(m)}(a>a_{Im}) \equiv {\hat{\Psi}}_{I-}(a),
\end{eqnarray}
which are similar to the Linde's wave function. $\Psi_{I}^{(m)}$ behaves as 
$\Psi_L$ in the range of $\delta <a <a_T$. But in the classical region, 
i.e. $a_T <a < a_{Im}$, there are differences to $\Psi_L$.
We call $\Psi_{I}^{(m)}$ the modified Linde wave function.  \\
Now, we are interested in the form of our wave function in the limiting
case of constant space dimension, i.e. $C \to +\infty$. In Appendix B, we
show that as $C \to +\infty$ the relevant integral behaves in the following
way:
\begin{equation}
\label{IV.78}
\lim_{C \to +\infty}{\int_{a_T}^{a_{Im}} p_I(a')da'}
=\lim_{C \to +\infty}{\frac{\pi {a_0}^3 e^{2C}}{2G^{3/2}}} \to \, +\infty.
\end{equation}
Substituting this limiting behavior in (\ref{IV.68}) and (\ref{IV.70}),
we see that the relevant terms are not well--defined. But we may assume cases 
in which  $C$ accepts those values corresponding to the relation 
(\ref{IV.74}) or (\ref{IV.74b}). 
It can easily be seen that in this limit 
$\tilde \Psi_{I-}(a)$, $\Psi_{I+}(a)$ and $\Psi_{I-}(a)$ tend to 
$\tilde \Psi_{-}(a)$, $\Psi_{+}(a)$, $\Psi_{-}(a)$, respectively.
Now, taking the limit of $C$ to infinity corresponding to $n \to +\infty$,
our wave functions approach to the Hartle--Hawking one:
\begin{eqnarray}
\lim_{n \to +\infty}\Psi_I^{(n)}(\delta<a<a_T)\equiv \Psi_{HH}(0<a<l_{\Lambda}),\\
\lim_{n \to +\infty}\Psi_I^{(n)}(a_T<a<a_{Im})\equiv \Psi_{HH}(a>l_{\Lambda}).
\end{eqnarray}
If the limiting behavior of $C$ in the relation (\ref{IV.74b}) corresponds to
$m \to +\infty$, then we obtain the modified Linde wave function:          
\begin{eqnarray}
\lim_{m \to +\infty}\Psi_I^{(m)}(\delta<a<a_T)\equiv \Psi_{L}(0<a<l_{\Lambda}),\\
\lim_{m \to +\infty}\Psi_I^{(m)}(a_T<a<a_{Im})\equiv 
i\,[\Psi_{+}(a)-\Psi_{-}(a)].
\end{eqnarray}
It should be noted that in this limiting case, the second barrier in the large
scale factors is removed and our potential $U_I(a)$ tends to
the standard potential $U(a)$.
Similarly, we can show that the same is true for the wave function
$\Psi_{II}(a)$. \\
We conclude that assuming the boundary condition (\ref{IV.63c}) and using
WKB approach we always have contracting and expanding terms in the classicl
region, i.e. Vilenkin wave function can not be derived from our general
wave function. This is due to the form of our potential which has a second
barrier at large scale factors, independent of the $C$--value. Therefore, we
may say that the Vilenkin wave function is structurally unstable with
respect to the variation of dimension.

\subsection{The probability density} 
\label{subsec-tunneling}

It is interesting to find the probability density in our model universe and
compare it to its value of the de Sitter minisuperspace in 3--space. 
From Eq.(\ref{V.8}), we can 
calculate the probability density 
\begin{equation}
\label{IV.200}
{\cal{P}}_I \sim \exp(-2 \int_{\delta}^{a_T} |p_I(a')| da'),
\end{equation}
for the potential barrier of $U_I(a)$ in the region $\delta < a <a_T$.
As before, we take $\Lambda=3G^{-1}$.
The integral can be calculated numerically for different values of $C$:
\begin{eqnarray}
\label{IV.201}
{\cal{P}}_I \sim \cases {\exp(-2.72)\, & if $C=1258800$, \cr
                \exp(-2.28 \times 10^{-60})\, & if $C=1678.8$. \cr}
\end{eqnarray}
The probability density decreases as $C$ increases, i.e. as the 
height of the potential barrier increases. 
Consequently, in the limit $C \to +\infty$, 
corresponding to the constant space dimension, the probability density has 
its minimum value given by Eq.(\ref{V.8}):
\begin{equation}
\label{IV.202}
\lim_{C \to +\infty}{{\cal{P}}_{I}} 
={\cal {P}} \sim \exp(-\frac{3 \pi}{G \Lambda}) =\exp(-\pi).
\end{equation}
This is the same as the tunnelling probability proposed by Vilenkin and
Linde. There is recently a controversy between Hartle, Hawking, and Turok
on one side and Vilenkin and Linde on the other side on the sign of the 
action in the probabity density (\cite{33}--\cite{38}, \cite{44}).
Our calculation of the probability
leads to the proposal of Vilenkin and Linde.
There is another barrier in the region $a_{Im}<a<+\infty$ for the potential
$U_I$, as seen by Fig. 7. The probability density for this potential barrier: 
\begin{equation}
{\hat{{\cal{P}}}}_{I} \sim \exp(-2 \int_{a_{Im}}^{+\infty} |p_{I}(a')| da'),
\end{equation}
is less than the probability density of the potential barrier in small scale 
factor. This is due to the height of the barrier for large scale factor 
is much more than the height of the barrier in small scale factor, 
see Fig. 7. The probability density 
\begin{equation}
\label{IV.203}
{\cal{P}}_{II} \sim \exp(-2 \int_{\delta}^{a_T} |p_{II}(a')| da'),
\end{equation}
for the potential barrier of $U_{II}(a)$ in the region $\delta < a < a_T$ 
and 
\begin{equation}
{\hat{{\cal{P}}}}_{II} \sim \exp(-2 \int_{a_{IIm}}^{+\infty} |p_{II}(a')| da'),
\end{equation}
for the potential barrier in the region $a_{IIm}<a<+\infty$
can similarly be
calculated and leads to similar results like $U_{I}(a)$.
The existence of this barrier for large scale factors 
and the behavior of the potential in that region has
cosmological consequences which will be discussed in 
a forthcoming paper \cite{rman}. \\
Assuming $\Lambda=3G^{-1} \sim 10^{38} GeV^2$, the Euclidean action 
$S_E=(-\frac{3 \pi}{G \Lambda})$ takes the value $(-\pi)$. This large value 
for $\Lambda$--term is estimated in modern theories of elementary particle.
Astronomical observations indicates that the cosmological constant is much 
less than this value. Assuming $\Lambda \leq 10^{-80}GeV^2$, the 
Euclidean action is very large and negative ($S_E \sim -10^{120}$ typically).
This gives the probability density proportional to 
${\cal{P}} \propto \exp(-10^{120})$. Therefore, the probability density in the
de Sitter minisuperspace depends on the assumed value for the cosmological
constant. From the above discussion we conclude that for a given value of the
cosmological constant, the probability density in our model is much more than 
the probability density of the de Sitter minisuperspace in 3--space. 
Therefore, the creation of a closed Friedman universe with
variable spatial dimension is much more probable than the creation of a 
closed de Sitter universe in 3--space. \\
So in our model, the universe tunnels from the first turning point, 
$a= \delta$ to the second turning point, $a=a_T$, expands up to the third 
turning point $a=a_{Im}$ (or $a_{IIm}$ for $U_{II}$). It may then starts 
tunneling through the potential barrier in the region $a>a_{Im}$ (or $a_{IIm}$).
Seeing the process classically, the point $a_{Im}$ (or $a_{IIm}$) may act as 
a turning point at which the universe begins to contract \cite{41}.\\

\section{CONCLUDING REMARKS}
\label{sec-concluding}
The idea of the space dimension to be other than three is relatively old,
at least since the work of Kaluza and Klein. But it seems that the
variability of space dimension is a relative new idea. We accept it as
a viable alternative and formulate a cosmological toy model based on a
Lagrangian to look for its consequences. 
In our formulation, there is a constraint and a scalar parameter,
$C$, accepting every positive value.
The limiting case of
$C \to +\infty$ corresponds to the standard model of constant space dimension.
Taking the present spatial dimension, $D_0$, to be equal to 3, and assuming
the space dimension at the Planck length, $D_{Pl}$, to be about 3, or one of
the values of 4, 10, or 25 coming from the superstring theories, the
corresponding values of $C$ are of the order $10^6, 10^3, 10^2, 10^2$,
respectively. For $D_0=2$ as a fractal dimension for matter distribution in
the universe, coming from cosmological considerations, we obtain different
values of $C$ corresponding to $D_{Pl}=$ 3, 4, 10 or 25.\\
After a critical review  of the previous works on the decrumpling universe
model, we generalized the Hawking--Ellis action for the prefect fluid universe
in 3--space to the case of variable space dimension. Using this generalized
action for a perfect fluid, we have then formulated the action for a FRW
universe filled by a prefect fluid and allowing the space dimension to be
a dynamical scalar variable. We have seen that this generalization is not
unique. In contrast to the earlier works, we have taken into account the
dependence of the measure of the integral on the spatial dimension. The
Lagrangian and the equations of motion and also the time evolution equation
of the spatial dimension are then obtained.\\
Using the time evolution equation of the spatial dimension, we have obtained
the classical turning points of our model universe. It turns out that 
the space dimension could have been up to infinity at the beginning of the
expansion phase of the universe, where the universe has a minimum size
equal to $\delta$. This is the only singularity we encounter in our model.\\ 
We then turn to quantum formulation of our toy model and write down the
Wheeler--De Witt equation for our universe. Imposing the appropriate
boundary condition at $a \to +\infty$ and using the semicalssical
approximation we have obtained the wave function of our model, which is
real but more general than its standard counterparts. As a result, the
Hartle--Hawking wave function is obtained for special values of the constant
$C$. Other values of $C$ lead to a modified Linde wave function. In the
standard limit of constant space dimension we may have a general, 
Hartle--Hawking, or a modified Linde wave function, but that of Vilenkin is
not obtained. This is due to the special shape of the potential which has
always a second barrier for large scale factors and only vanishes in the
limiting case of $C \to +\infty$. We have therefore concluded that the
Vilenkin wave function is, in respect of varing the space dimension,
structurally unstable.\\
Finally we have calculated the probability density for the potential barrier
in small scale factor. In the limit of constant spatial dimension, this
probability density approaches to its corresponding value of the de Sitter
minisuperspace as suggested by Vilenkin and Linde. In general, the probabilty
density for our toy model is larger than that of the standard quantum
cosmology.\\
Our toy model shows that it is possible to formulate a Lagrangian for a
cosmological model with dynamical space dimension. It removes the usual
singularities in physical quantities like scale factor of the universe
because the minimum scale factor of the universe is $\delta$.
It gives us a model of how a probable space dimension of 10 or 25 at the
Planck length can be incorporated in the usual picture of $(1+3)$--dimensional
space time and what its consequences are. We are currently studying
other cosmological consequences of the model, specially those which can be
compared to obsevational data.

\section*{Acknowledgments}  

One of us (F.N) is grateful to Professor Alex Vilenkin and Professor 
Manuel Bronstein for helpful discussions.

\subsection*{APPENDIX A: Volume of spacelike sections}

The D--dimensional space metric is defined as
(see Eq.~\ref{III.1})
\begin{eqnarray}
d{\Sigma_k}^2&=&d\chi^2+F^2(\chi)[{d\theta_1}^2
+\sin^2\theta_1{d\theta_2}^2+...\nonumber\\
&+&\sin^2{\theta}_1...\sin^2\theta_{D-2}d\theta_{D-1}^2],
\end{eqnarray}
where
$$
0\leq \theta_i \leq \pi\;\;\;\;\;\;\;i=1,...,D-2\;\;\;\;\;\;0\leq \theta_{D-1}
\leq 2\pi.
$$
The radius $F(\chi)$ is expressed as
\begin{eqnarray}
F (\chi) = \cases {\sin {\chi}\, & if $k=+1, \;0\leq \chi \leq \pi$, \cr
                \chi\, & if $k=0,\;0 \leq {\chi} < \infty$, \cr
                \sinh\, \chi & if $k=-1,\;0 \leq {\chi} < \infty$. \cr}
\end{eqnarray}
We know consider the different cases $k=0, \pm 1$, separately:\\
i) $k=+1$:\\
$V_D$ is the volume of a unit D--sphere, $S^D$:
\begin{eqnarray}
V_D&=& \int_0^\pi \sin^{D-1} \chi d \chi \int_0^{2 \pi} d \theta_{D-1}\nonumber\\ 
&\times&\prod_{m=2}^{D-1, D>2} \int_0^{\pi} d \theta_{D-m} \sin^{m-1} \theta_{D-m}\nonumber\\
&=&\frac{2 \pi^{(\frac{D+1}{2})}}{\Gamma(\frac{D+1}{2})}.
\end{eqnarray}
ii) $k=0$:\\
In this case $V_D$ is infinite. Introducing a cut-off, $\chi_c$,
as a very large number, $V_D$ has the form
\begin{eqnarray}
V_D&=& \int_0^{\chi_c} {\chi}^{D-1} d\chi \int_0^ {2 \pi} d \theta_{D-1} \nonumber\\
&\times& \prod_{m=2}^{D-1,D>2} \int_0^{\pi} d\theta_{D-m} \sin^{m-1} \theta_{D-m}\nonumber\\
&=&\frac{ {\chi_c}^D \pi^{D/2}}{\Gamma(\frac{D}{2}+1)}.
\end{eqnarray}
iii) $k=-1$:\\ 
Again, using the case of $\chi_C$, we obtain 
\begin{eqnarray}
V_D&=& \int_0^{\chi_c} {\sinh}^{D-1} \chi d \chi \int_0^{2 \pi} d \theta_{D-1}\nonumber\\
&\times&\prod_{m=2}^{D-1, D>2} \int_0^ {\pi} d \theta_{D-m} \sin^{m-1} \theta_{D-m}\nonumber\\
&=&\frac{ 2 \pi^{D/2}}{\Gamma(\frac{D}{2})}f(\chi_c).
\end{eqnarray}
where
\begin{equation}
f(\chi_c) = \cases {\frac{1}{2^{2m-1}} \sum_{k=0}^m (-1)^k{2m \choose k}
\frac{\sinh (2m-2k) \chi_c}{2m-2k}, 
& if $D=2m+1$, \cr \\
\frac{1}{2^{2(m-1)}} \sum_{k=0}^{m-1} (-1)^k{ 2m-1 \choose k}
\frac{\cosh (2m-1-2k) \chi_c -1}{2m-1-2k},
& if $D=2m$. \cr} 
\end{equation}
where $m$ is an integer number.
For details of the about integration see \cite{mat}. In general 
$\chi_C$ tends to infinity so that for $k=0, -1$, the volume of the spacelike 
sections are infinite.\\
It should be emphasized that the above expression 
for $V_D$ is valid for integer dimension.  
Volume of fractal structures having a non--integer dimension is beyond the 
scope of this paper.

\subsection*{APPENDIX B: Limiting behavior of our wave function for 
$C \to +\infty$}
Our goal is to obtain the behavior of integrals in the wave function
(\ref{IV.68}) and (\ref{IV.70}) as $C \to +\infty$:
$$
\lim_{C \to +\infty}{\int_{a_T}^{a_{Im}} p_I(a)da}.
$$
We calculate it by expanding $p_I(a)$ in inverse powers of $C$:
\begin{equation}
\label{IV.77}
p_I(a)=\frac{3 \pi a}{2G}\sqrt{(\frac{a^2}{G}-1)}+\frac{f(a)}{C}+
{\cal{O}}(\frac{1}{C^2})+...,
\end{equation}
where we take $D_0=3$ and 
\begin{equation}
\label{IV.78a}
f(a)=\frac{27 \pi a}{4G}(\ln \frac{a}{a_0}) \{ \frac{5}{6}(-1+\frac{a^2}{G})^{-1/2}
-(-1+\frac{a^2}{G})^{1/2} [\frac{29}{6}+\ln \pi -\psi(2)] \},
\end{equation}
where $\psi(2) \simeq 0.42$. 
From Table III, we know that for $C \sim 10^6$ or so,
the value of $a_T$ is of the order of the Planck length.
Therefore, we can also expand $a_T$ in inverse powers of $C$: 
\begin{equation}
\label{IV.79}
a_T=\sqrt G +\frac{A}{C}+{\cal{O}}(\frac{1}{C^2})+...,
\end{equation}
where $A$ is a constant. From Eqs.(\ref{IV.77})--(\ref{IV.79}) we obtain
\begin{equation}
\label{IV.80}
\int^{a_{Im}}_{a_T} p_I(a) da=\frac{\pi}{2}(-1+\frac{{a_{Im}}^2}{G})^{3/2}+
\frac{A_1}{C}+{\cal{O}}(\frac{1}{C^2})+...,
\end{equation}
where $A_1$ is a constant. As previously mentioned, 
the relation between $a_{Im}$ and $C$ is
\begin{eqnarray}
\label{IV.81}
a_{Im}=a_0 e^{\frac{2C}{3}}.
\end{eqnarray}
Here $a_0$ is the scale factor corresponding to $D=D_0$. 
Inserting (\ref{IV.81}) into (\ref{IV.80}), we are led to 
\begin{equation}
\label{IV.82}
\lim_{C \to +\infty}{\int_{a_T}^{a_{Im}} p_I(a)da}=
\lim_{C \to +\infty}{{\frac{\pi}{2}}(-1+\frac{{a_0}^2 
e^{\frac{4C}{3}}}{G})^{3/2}} \sim 
\lim_{C \to +\infty}{\frac{\pi {a_0}^3 e^{2C}}{2G^{3/2}}}.
\end{equation}

\newpage

\subsection*{FIGURE CAPTIONS}

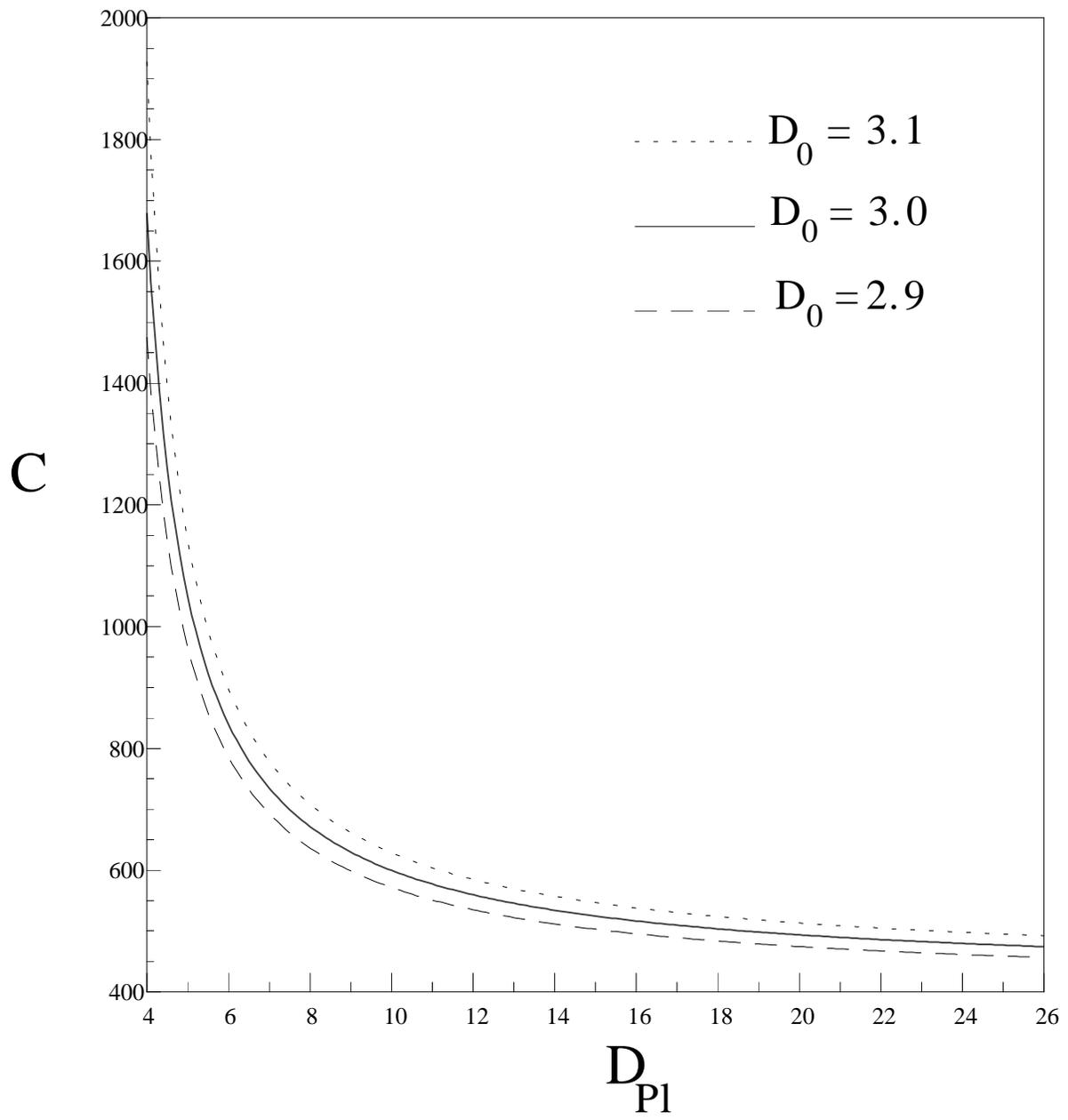
FIG. 1. The values of $C$ as a function of $D_{Pl}$ when $4 < D_{Pl} < 26$ and
$D_0$ taking the values 2.9 (dashed line), 3.0 (solid line),
3.1 (dotted line). Using Eq.(\ref{c3}) it is easy to show that if we take 
$D_0=3$ for $D_{Pl}=3$ and $D_{Pl} \to +\infty$, we have $C \to +\infty$ and 
$C=D_0 \ln \frac{{H_0}^{-1}}{l_{Pl}} \simeq 419.7,$ respectively.

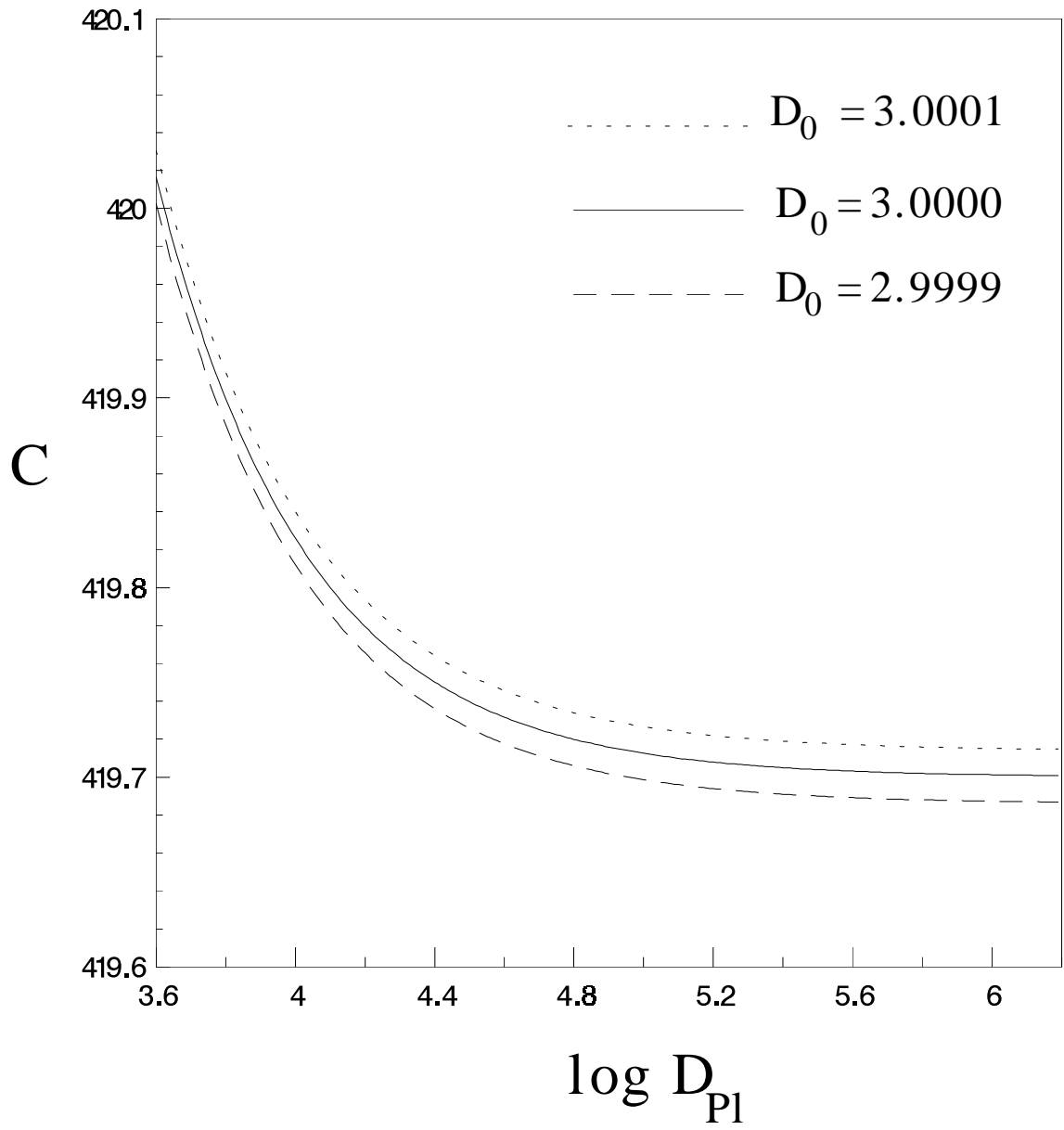
FIG. 2. The values of $C$ as a function of $\log D_{Pl}$ when 
$3.6<\log D_{Pl}<6.2$ and $D_0$ taking the values 2.9999 (dashed line),
3.0000 (solid line), 3.0001 (dotted line).

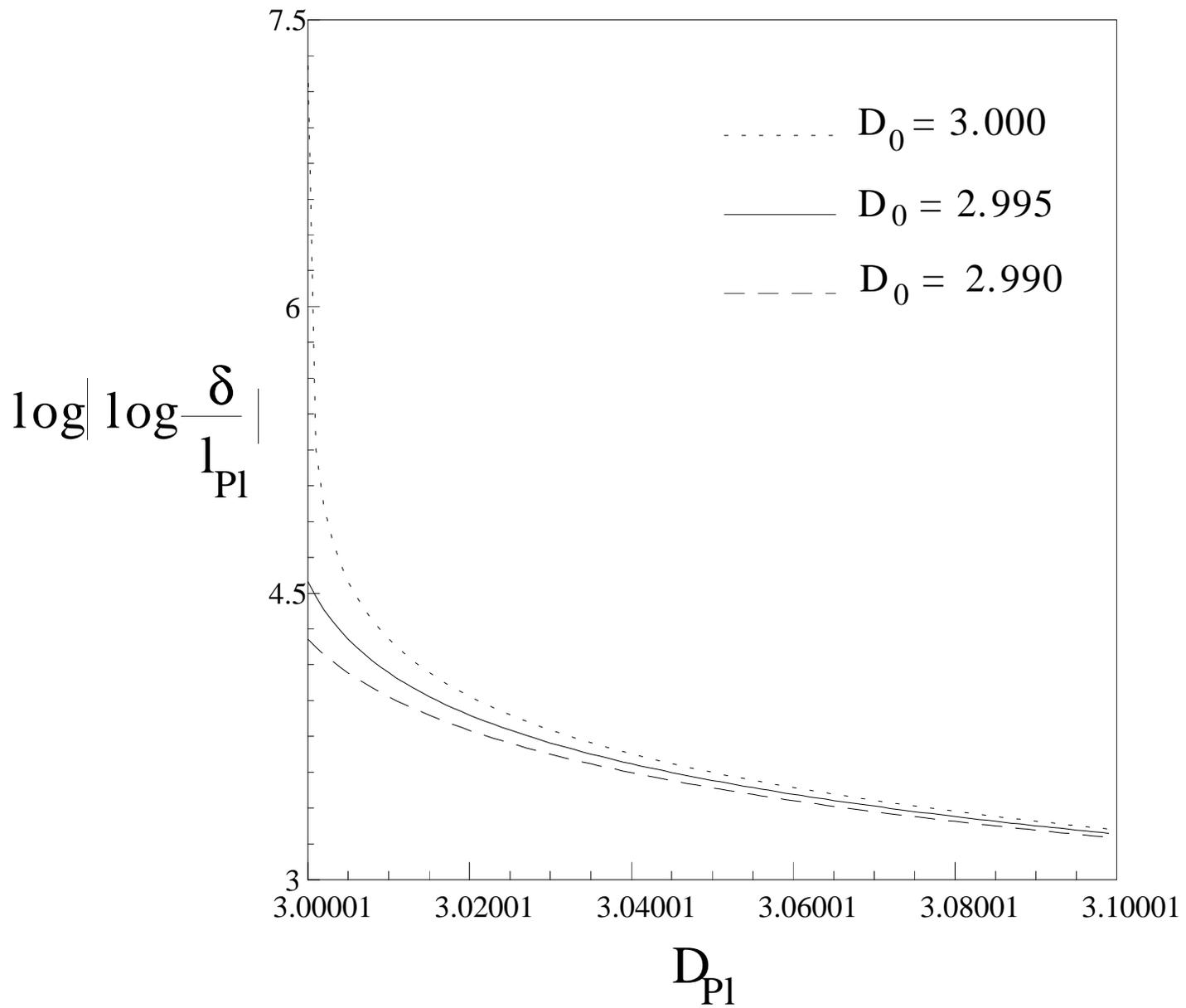
FIG. 3. The values of $\log|\log \frac{\delta}{l_{Pl}}|$ as a function of 
$D_{Pl}$ when $D_{Pl} \simeq 3$ and $D_0$ taking the values 2.990 
(dashed line), 2.995 (solid line), 3.000 (dotted line).

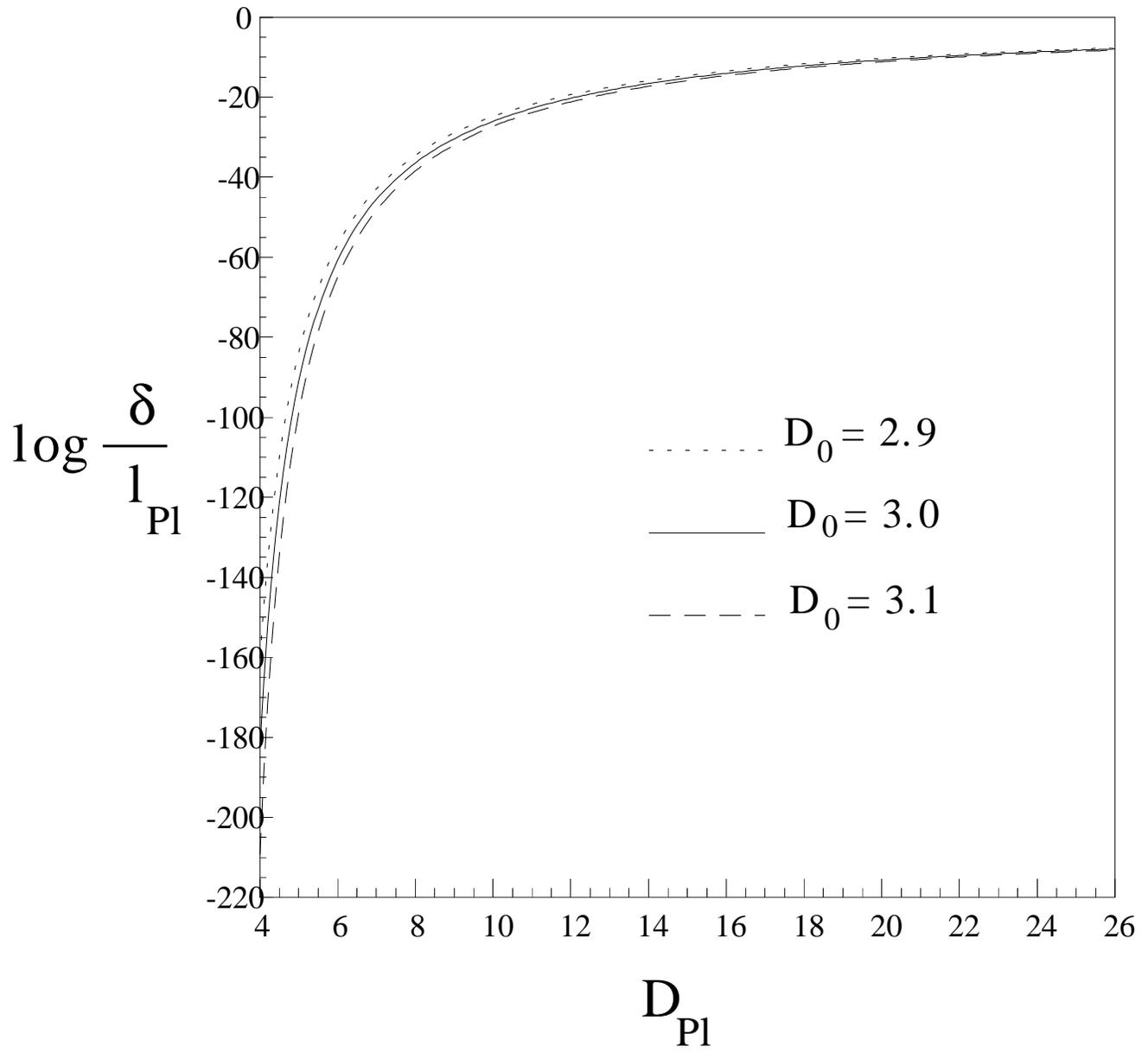
FIG. 4. The values of $\log (\frac{\delta}{l_{Pl}})$ as a function of 
$D_{Pl}$ when $4<D_{Pl}<26$ and $D_0$ taking the values 3.1 (dashed line), 
3.0 (solid line), 2.9 (dotted line). Using Eq.(\ref{c4}), it is easy to show 
that if we take $D_0=3$, for $D_{Pl}=3$ and $D_{Pl} \to +\infty$, we have 
$\log \frac {\delta}{l_{Pl}} \to -\infty$ and $\log \frac{\delta}{l_{Pl}}=0$,
respectively.

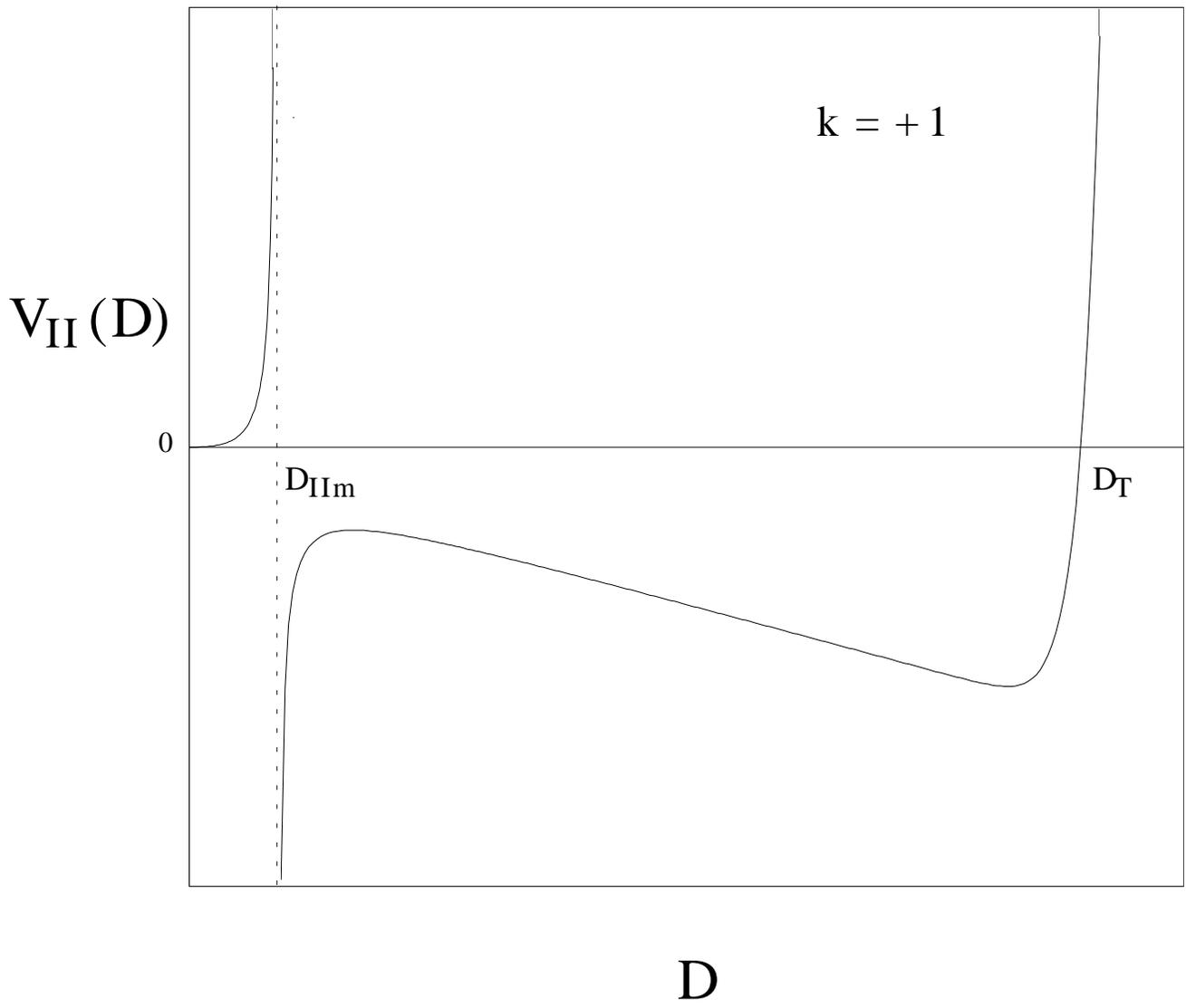
FIG. 5. The generic shape of the potential $V_{II}(D)$ for $k=+1$. The
kinetic energy, ${\dot D}^2$, is positive in the classical region,
$D_{IIm}<D<D_T$. The other values of $D$ are corresponding to non--classical 
region in which the kinetic energy is negative.

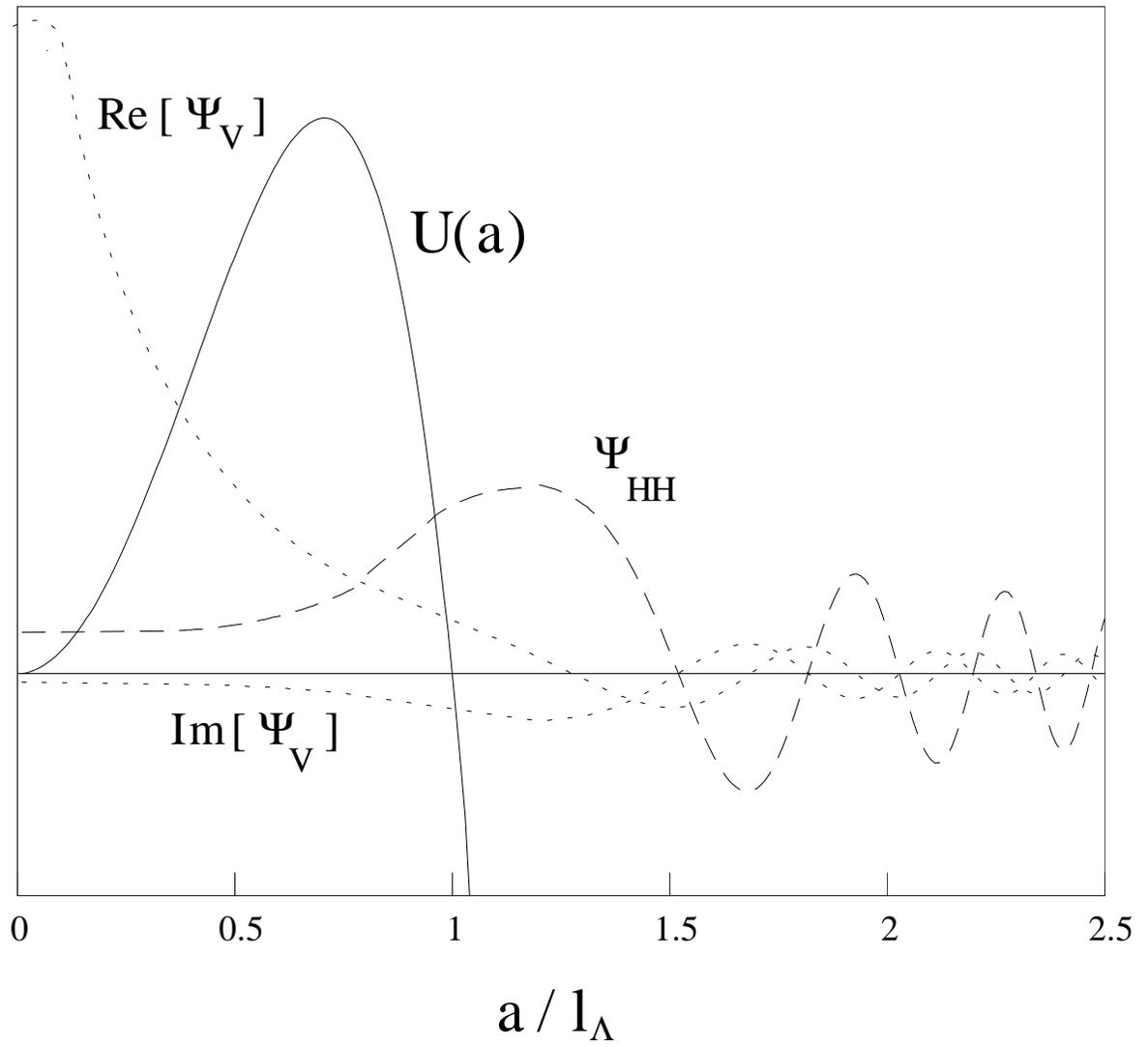
FIG. 6. The potential $U(a)$ is shown by a solid line. The wave function
of Hartle--Hawking (dashed curve) and that of Vilenkin (dotted curve)
are shown. 
The real and imaginary parts of Vilenkin tunneling wave function are 
so indicated. The Hartle--Hawking wave function is real.

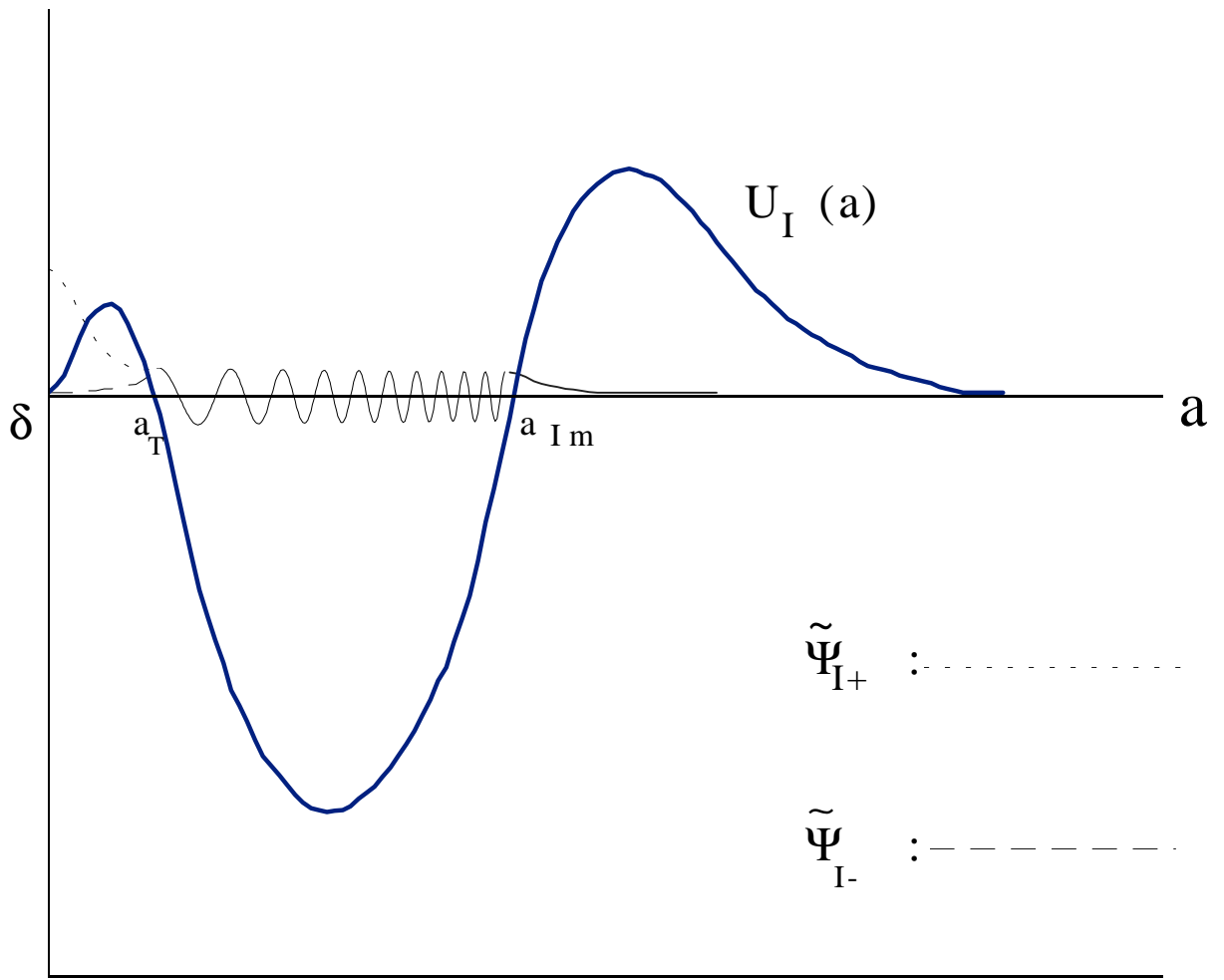
FIG. 7. The generic shape of the potential $U_I(a)$.
The growing wave function ${\tilde \Psi}_{I-}$ (dashed line) and the 
decreasing wave function ${\tilde \Psi}_{I+}$ (dotted line) are shown.  
The under--barrier wave function, in the region $\delta<a<a_T$ is the sum of 
${\tilde \Psi}_{I-}$ and ${\tilde \Psi}_{I+}$ (see Eq.\ref{IV.70}).
In the region $a_T<a<a_{Im}$ there is an oscillating wave function.
For $a>a_{Im}$ the decreasing wave function ${\hat \Psi}_{I-}$ is also
indicated. This shape is drawn approximately and it shows only the behavior 
of the potential $U_I(a)$. The similar shape can be drawn for the potential 
$U_{II}(a)$ and its corresponding wave function, except for $a_{IIm}$ instead 
of $a_{Im}$.


\begin{thebibliography}{99}
\bibitem{44}Vilenkin in his first paper on quantum cosmology 
\cite{{avil}, {43}} used the standard Wick rotation for Euclideanization of  
the Lorentzian path integral of the wave function, so that the probability
density in his proposal gave the same sign of the exponent as the 
no--boundary proposal. But later he realized \cite{avilen} that it did not 
correspond to the picture of ``tunneling from nothing'' which he had in mind 
from the beginning.
\end{thebibliography}
\end{document}